\documentclass{article}
\usepackage[english]{babel}

\usepackage[letterpaper,top=2cm,bottom=2cm,left=3cm,right=3cm,marginparwidth=1.75cm]{geometry}

\usepackage{amsmath}
\usepackage{amssymb}
\usepackage{graphicx}
\usepackage[colorlinks=true, allcolors=blue]{hyperref}
\usepackage{natbib}
\title{Diffusion-based subsurface CO$_2$ multiphysics monitoring and forecasting}

\author{Xinquan Huang$^{1,2}$, Fu Wang$^{1}$, Tariq Alkhalifah$^{1}$\\
$^{1}$King Abdullah University of Science and Technology\\
$^{2}$University of Pennsylvania\\
\texttt{\{xinquan.huang,fu.wang,tariq.alkhalifah\}@kaust.edu.sa}\\}
\date{}
\begin{document}
\maketitle

\begin{abstract}
Carbon capture and storage (CCS) plays a crucial role in mitigating greenhouse gas emissions, particularly from industrial outputs. 
Using seismic monitoring can aid in an accurate and robust monitoring system to ensure the effectiveness of CCS and mitigate associated risks. 
However, conventional seismic wave equation-based approaches are computationally demanding, which hinders real-time applications. 
In addition to efficiency, forecasting and uncertainty analysis are not easy to handle using such numerical-simulation-based approaches.
To this end, we propose a novel subsurface multiphysics monitoring and forecasting framework utilizing video diffusion models. 
This approach can generate high-quality representations of CO$2$ evolution and associated changes in subsurface elastic properties. 
With reconstruction guidance, forecasting and inversion can
be achieved conditioned on historical frames and/or observational data.
Meanwhile, due to the generative nature of the approach, we can quantify uncertainty in the prediction.
Tests based on the Compass model show that the proposed method successfully captured the inherently complex physical phenomena associated with CO$_2$ monitoring, and it can predict and invert the subsurface elastic properties and CO$_2$ saturation with consistency in their evolution.
\end{abstract}

\section*{Key points}
\begin{itemize}
    \item We develop a novel monitoring and forecasting framework using video diffusion models of subsurface changes due to, e.g., CO$_2$ injection.
    \item This approach can be a simulation engine for multiphysics evolution and maintain physical consistency between multiple variables.
    \item The proposed method is applied to multiphysics monitoring and forecasting for permeability, CO$_2$ saturation, velocity, density, and RTM images.
\end{itemize}

\section{Introduction}
Carbon capture and storage (CCS) play a crucial role in mitigating greenhouse gas emissions, particularly from industrial outputs. This significance is underscored in the recent International Panel on Climate Change report \citep{ipcc_climate_2022}. 
CCS involves capturing carbon dioxide (CO$_2$), transporting it, and injecting it into subsurface storage sites. 
However, to ensure its effectiveness and mitigate associated risks, monitoring, verification, and accounting of the injected CO$_2$ plume are essential \citep{sun_denoising_2023}. 
Effective monitoring hinges on accurate and robust systems. 
Seismic monitoring, as highlighted by \cite{lumley_4d_2010}, can provide such a system that is both accurate and covers a large area. 
It operates on the principle that CO$_2$ injection alters the elastic properties of the subsurface, such as velocities and densities. 
These changes can be inferred, to some extent, from the imaging and inversion of seismic recordings \citep{bosch_seismic_2010,hicks_time-lapse_2016,li_target-oriented_2022,hu_feasibility_2023}. 
However, conventional methods like statistical (e.g., Monte Carlo simulation) or wave equation-based approaches are computationally demanding and thus costly for realistic applications. 
In addition to the monitoring, predicting the future states of CO$_2$ flow is also important as they can provide information on how much CO$_2$ can be stored safely and help us optimize our decisions with respect to injections.
Using the above conventional methods, we struggle to track the CO$_2$ plume evolution and predict future subsurface states.
Solving non-linear partial differential equations for a multi-phase flow phenomenon \citep{pruess_numerical_2011,ringrose_co2_2020} can help us predict future states based on historical states.
Although accurate and generic, this prediction is still computationally expensive because of the formulations involved in the iterative solvers \citep{blunt_simulation_1992,graupner_coupled_2011}, especially for the inversion application involved with a repeated simulation \citep{witte_fast_2023}.  

In the era of artificial intelligence, modern machine-learning techniques have opened new avenues, data-driven approaches, to tackle the above issues. 
Recent studies have leveraged machine learning techniques for various aspects of CCS monitoring and forecasting. 
\cite{li_prediction_2018} proposed the use of an artificial neural network to predict the probability of long-term leakage of wells in CO$_2$ sequestration operations. 
\cite{zhou_data-driven_2019} proposed to use a spatial-temporal densely connected convolutional neural networks to detect CO$_2$ leakage from seismic data, where the network design is geared to capture spatial and temporal characteristics of seismic data.
Beyond leakage detection, \cite{zhong_predicting_2019} proposed a conditional deep convolution generative adversarial network as a surrogate model to predict CO$_2$ plume migration. \cite{stepien_continuous_2023} enhanced this approach by predicting CO$_2$ plume migration with observations at three well locations.
\cite{wen_ccsnet_2021} proposed the CCSNet to efficiently handle an entire class of multiphase flow problems, providing predictions of CO$_2$ gas saturation distribution, pressure buildups, the molar fractions of CO$_2$ and fluid densities for gas and liquid phases, in CCS modeling. \cite{wen_accelerating_2022} further accelerate the machine learning-based prediction using Fourier neural operators \citep{li_fourier_2020}.
\cite{witte_fast_2023} applied wavelet neural operators to enable fast CO$_2$ saturation simulation on large-scale geomodels. 
To incorporate more observation to better monitor the CO$_2$ migration and optimize reservoir management, \cite{li_coupled_2020} proposed a coupled time-lapse full waveform inversion framework to estimate subsurface properties such as rock permeability and porosity from seismic data by intrusive automatic differentiation.
\cite{louboutin_learned_2023} further expanded this cascaded multiphysics inversion with machine learning priors, which represent the permeability using normalizing flow \citep{rezende_variational_2015}.
Based on this cascaded multiphysics framework, \cite{yin_derisking_2023} developed an explainable leakage detection network from time-lapse seismic data.
Beyond cascaded multiphysics inversion, \cite{sun_denoising_2023} directly learn the mapping from seismic shot gathers to the CO$_2$ plume using denoising diffusion probabilistic models.
\cite{gahlot_inference_2023} proposed to use conditional normalizing flows to obtain the CO$_2$ saturation based on the available seismic images and well log observations.
\cite{chen_co2seg_2024} proposed to treat the CO$_2$ monitoring task as a segmentation task to save the interpretation time.
Yet, these methods primarily focus on leakage detection and CO$_2$ plume interpretation, accelerating CO$_2$ plume simulation and CCS modeling, single physics and 2D mapping with generative models, or a cascaded multiphysics inversion. 
Real-world monitoring and forecasting of subsurface CO$_2$ injections involve a complex multiphysics process encompassing fluid flow simulation, rock physics, and wave propagation.

To achieve an end-to-end multiphysics inversion and maintain the consistency of the evolution of different variables in terms of physical laws, we introduce a novel generative model-based framework for monitoring and forecasting in this paper. 
Rather than directly performing the deterministic prediction or the cascade inversion, we train the model to represent the joint distribution among multiple physical variables and perform the monitoring and forecasting with conditional sampling in this distribution.
Specifically, we chose denoising diffusion probabilistic models (DDPMs) \citep{ho_denoising_2020} to represent the data distribution.
Two-dimensional DDPMs have shown promising results in seismic processing \citep{durall_deep_2023}, diffusion regularized full waveform inversion \citep{wang_prior_2023}, the inverse mapping between the seismic shot gathers and the CO$_2$ plume shape \citep{sun_denoising_2023}, controllable velocity synthesis \citep{wang_controllable_2024}, and seismic diffraction separation and imaging \citep{zhang_conditional_2024}.
Instead of a 2D version, our framework trains a diffusion model to learn the data distribution of 3D multiphysics evolution (spanning 2D along the spatial domain and 1D along time-lapse direction). 
Those 3D volumes incorporate variables such as permeability, CO$_2$ saturation, velocity, density, and subsurface seismic imaging results using reverse time migration (RTM). 
The philosophy behind the use of 3D volume (video) prediction is that the physical process in the real world can be regarded as video sequences. 
Thus, the forecasting and inversion process will be thought of as the tasks for video generation given the history frames or partial information. 
Unlike recursively predicting the future states along the time axis using a neural operator, which might have errors accumulation \citep{ding_diffusion_2024}, the video diffusion model can simultaneously generate multistep future states without the need for sequential inference.
During inference, we employ reconstruction guidance to facilitate forecasting and inversion using historical frames or observed RTM images. 
Thanks to the probabilistic nature of the diffusion model, we can easily perform uncertainty analysis for forecasting and inversion.
In this paper, we validate our method using the models extracted from the 3D Compass model \citep{jones_building_2012}.

The main contributions of this study include:
\begin{itemize}
    \item We propose a novel unified monitoring and forecasting framework using video diffusion models, and it inherently supports uncertainty analysis.
    \item We show that the video generation framework can be a simulation engine for multiphysics evolution, as well as maintain a physical consistency between different physical parameters, thus enabling subsequent forecasting and monitoring.
    \item We validate the proposed framework on the compass model as we demonstrate the effectiveness of the proposed method in handling multiphysics monitoring and forecasting, specifically for permeability, CO$_2$ saturation, velocity, density, and reverse time migration (RTM) images.
\end{itemize}

The remainder of the paper is organized as follows. In section \ref{sec:method}, we introduce the video diffusion models and the proposed framework for forecasting and monitoring based on video diffusion models with reconstruction guidance. Then, we share our numerical simulation for the data generation. In section~\ref{sec:exp}, we present numerical tests of the proposed method on compass models, including the results of unconditional generation, monitoring, forecasting, and uncertainty analysis. Finally, we discuss and summarize the work, and present insights in sections~\ref{sec:discuss} and~\ref{sec:conclusion}. 

\section{Methodology}
\label{sec:method}
Our proposed forecasting and inversion framework is based on video diffusion models. In this section, we begin with the definition of the forecasting and inversion frameworks as probabilistic predictions, which are smoothly connected to the denoising diffusion probabilistic models (DDPMs). 
After that, a detailed introduction about the key concepts and the pipeline behind the DDPMs as well as, the post-training controlled generation using reconstruction guidance will be provided.
Finally, we introduce the variables describing the reservoir state used in this study to validate the effectiveness of the proposed method and related multiphysics processes.

\subsection{Task definition}
\label{sec:method_task}
Both the forecasting and inversion can be formulated in a general form, in which, given the observations $\mathbf{y}$, which can be both the history states and current observation, we estimate multiple physical parameters $\mathbf{x}$ (e.g., velocity or density change, CO$_2$ saturation) corresponding to current or future states.
In a deterministic framework, it can be written as follows:
\begin{equation}
    \label{equ:determ}
    \mathbf{y} = f(\mathbf{x}),
\end{equation}
where $f$ denotes the modeling operator used for the inference of the future state given current/history states or for the inference of the current states as a response to observed data.
However, for CO$_2$ monitoring and related multiphysics evolution, due to the complexity of subsurface behavior of CO$_2$ injection and related physical processes, geological uncertainties (such as those in our porosity and permeability measurements), technical limitations of the current observations, which may not be able to fully capture small leaks of carbon dioxide or complex geological changes, data analytics and the choice of assumptions and parameters in the modeling process introduce uncertainty. In other words, it is challenging to obtain accurate monitoring and forecasting results. Uncertainty is unavoidable.

Thus, we recast Equation~\ref{equ:determ} to a form in which we can draw samples from the conditional probability distribution, $p(\mathbf{x}^{1:T}|\mathbf{y})$, over T-steps monitoring or forecasting conditional on observations or previous/historical states $\mathbf{y}$. 
For a proof of concept, we form $\mathbf{x}^t$ to represent 5 CO$_2$ related physical variables at time step $t$.
To model $p(\mathbf{x}^{1:T}|\mathbf{y})$, instead of using ensemble methods with a learned neural network \citep{fort_deep_2020} or using the Bayesian method \citep{wilson_bayesian_2022}, we use denoising diffusion probabilistic models (DDPMs) to model the distribution that allows us to draw samples from the distribution directly.

\subsection{Video Diffusion models} 
\begin{figure}
    \centering
    \includegraphics[width=\textwidth]{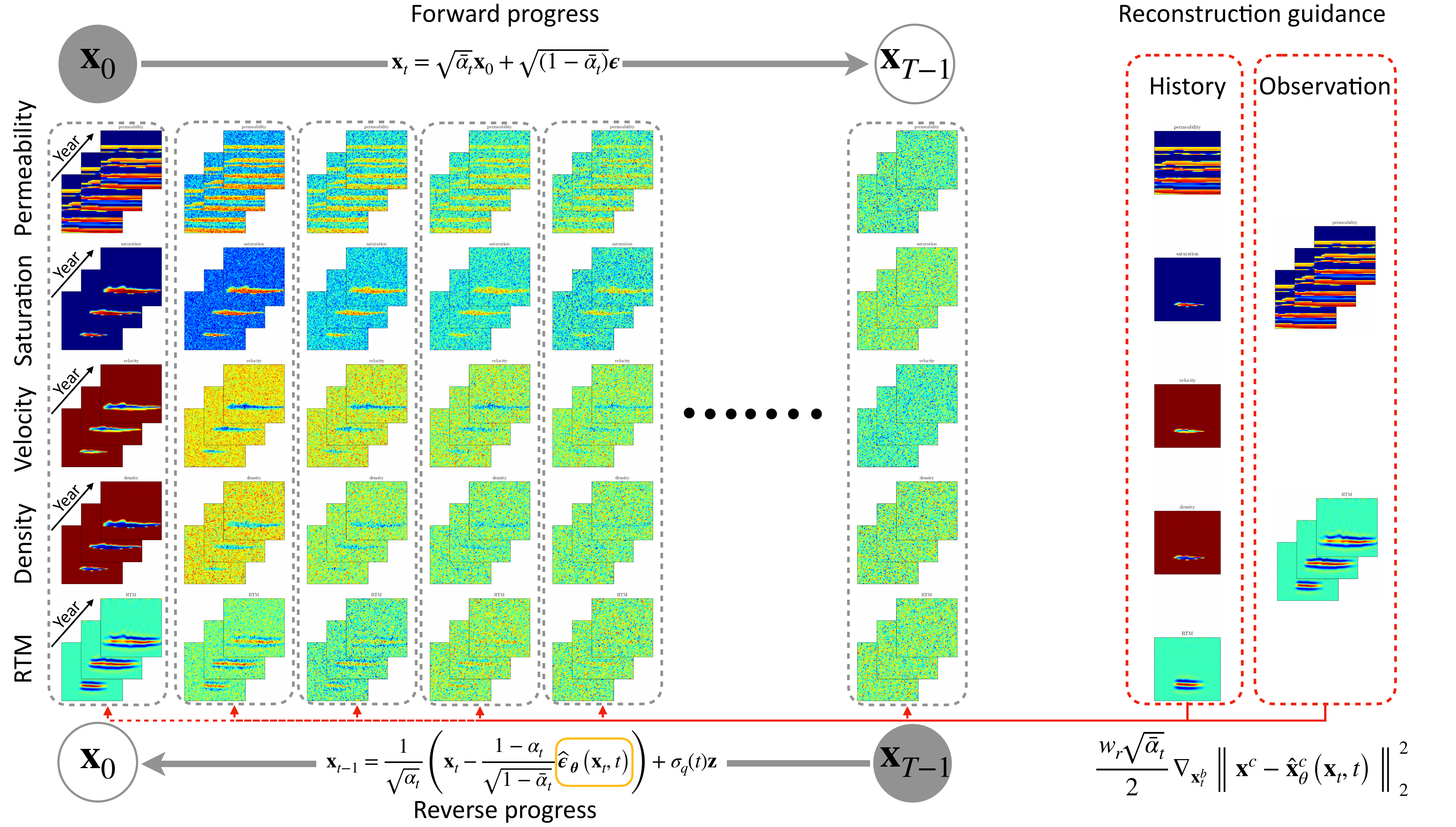}
    \caption{Workflow of the diffusion process and its reverse process, and the reconstruction guided generation for forecasting and inversion. The yellow box denotes the scaled learned score function. ``History" denotes the given conditions $\mathbf{x}^{0:t}$ for the forecasting tasks to predict $\mathbf{x}^{t:T}$, where superscript stands time-lapse frames. ``Observation" denotes partial physical variables of $\mathbf{x}^{0:T}$, e.g., RTM images, for the inversion of other variables.}
    \label{fig:diagram}
\end{figure}

Denoising diffusion probabilistic models (DDPMs) are a family of generative models that try to represent a data distribution which enables the generation of samples from the data distribution. 
They have shown a state-of-the-art image and video synthesis performance among other generative models, such as variational autoencoders \citep{kingma_auto-encoding_2013}, generative adversarial networks \citep{goodfellow_generative_2014}, normalizing flows \citep{rezende_variational_2015}.
The general idea behind DDPMs is to train a neural network to learn to transform a prescribed distribution (often a Gaussian distribution) to an approximation of the target data distribution.
The pipeline of the diffusion model includes two processes: training (a.k. forward process from $\mathbf{x}_0$ to $\mathbf{x}_{T-1}$) and sampling (a.k. reverse process from $\mathbf{x}_{T-1}$ to $\mathbf{x}_0$), shown in Figure~\ref{fig:diagram}. 
The subscript $T$ of the $\mathbf{x}_T$ here denotes the time step in the diffusion process.  
Our target is to fit the data distribution of spatial-temporal evolution for multiple variables, which are 3D volumes. 
We process it as videos and thus use video diffusion models \citep{ho_video_2022}.

During the training, we corrupt the video with noise by gradually adding a small amount of Gaussian noise, and then train a neural network as a denoiser to restore the clean videos.
For sampled data $\mathbf{x} \sim p(\mathbf{x})$, the corruption process (forward process) is a Gaussian process satisfying 
\begin{equation}
q\left(\mathbf{x}_t \mid \mathbf{x}_{t-1}\right)=\mathcal{N}\left(\mathbf{x}_t ;\sqrt{\alpha_t}  \mathbf{x}_{t-1}, (1-\alpha_t) \mathbf{I}\right), \quad 
q\left(\mathbf{x}_t \mid \mathbf{x}_0\right)=\mathcal{N}\left(\mathbf{x}_t ; \sqrt{\bar{\alpha}_t} \mathbf{x}_0, \left(1-\bar{\alpha}_t\right) \mathbf{I}\right),
\end{equation} 
where $q\left(\mathbf{x}_t \mid \mathbf{x}_{t-1}\right)$ denotes transition distribution, $\mathbf{x}_t$ denotes the corrupted videos at time-step $t$, where $0 \leq t \leq T-1$, $\alpha_t=1-\beta_t$, $\bar{\alpha}_t=\prod_{i=0}^t \alpha_i$, and $\beta_t$ is the noise scheduler tasked with scaling the noise so that the corrupted videos at time $T$ satisfy the Gaussian distribution $\mathcal{N}(0, \mathbf{I})$.
We chose the cosine noise scheduler in this paper as it provides better performance and quality \citep{nichol_improved_2021}. 
Similar to \citep{ho_denoising_2020}, instead of predicting the clean videos directly, the training process is realized by utilizing a neural network (3D Unet here) $\widehat{\boldsymbol{\epsilon}}_{\boldsymbol{\theta}}(\mathbf{x}_t, t)$ to estimate the Gaussian noise $\boldsymbol{\epsilon}$ for a given $\mathbf{x}_t$, and optimize the following loss function
\begin{equation}
    \label{equ:loss}
    \mathcal{L}=\mathbb{E}_{q\left(\mathbf{x}_t \mid \mathbf{x}_0\right)}\left\|\widehat{\boldsymbol{\epsilon}}_{\boldsymbol{\theta}}\left(\mathbf{x}_t, t\right)-\boldsymbol{\epsilon}\right\|^2,
\end{equation}
where one can rewrite $\mathbf{x}_t=\sqrt{\bar{\alpha_t}}\mathbf{x}_0+\sqrt{(1-\bar{\alpha_t})}\boldsymbol{\epsilon}$.

At the stage of sampling (inference), we draw random samples $\mathbf{x}_T$ from a Gaussian distribution and recursively denoise the videos by means of 
\begin{equation}
\label{equ:sample}
\mathbf{x}_{t-1}=\frac{1}{\sqrt{\alpha_t}}\left(\mathbf{x}_t-\frac{1-\alpha_t}{\sqrt{1-\bar{\alpha}_t}} \widehat{\boldsymbol{\epsilon}}_{\boldsymbol{\theta}}\left(\mathbf{x}_t,t\right)\right)+\sigma_q(t) \mathbf{z}
\end{equation}
until the final clean samples $\mathbf{x}_0$ are realized, where 
variance $\sigma_q(t)=\sqrt{\frac{\left(1-\alpha_t\right)\left(1-\bar{\alpha}_{t-1}\right)}{1-\bar{\alpha}_t}}$, $\mathbf{z}$ is the standard Gaussian noise. 

\subsection{Reconstrcution guided Conditional generation}
As mentioned in section~\ref{sec:method_task}, our goal is to draw samples from the conditional probability distribution, which is stored in the video diffusion model, conditioned by the observations to achieve forecasting, inversion, and probabilistic prediction.
Hence, to build $p(\mathbf{x}|\mathbf{y})$, the typical way is to directly incorporate the conditions in the neural network $\widehat{\boldsymbol{\epsilon}}_{\boldsymbol{\theta}}(\mathbf{x}_t, t)$ as an additional input to replace the neural network prediction in equation~\ref{equ:sample}.
This would implicitly modify the sampling trajectory in the diffusion process to correct the samples to not only satisfy the stored distribution of the data themselves, but also be consistent with the conditions \citep{ho_classier-free_2021}. 
However, this would require the retraining of the original unconditional diffusion model with sufficient data-label pairs to improve the generalization.
In addition, for multi-physics monitoring and forecasting promoted here, the conditions in our framework are the history/current states of multiple variables or those obtained from imaging current surface observations using, for example, reverse time migration (RTM), where such images can be represented as $\mathbf{x}^{\mathbf{c}}$. 
The superscript $\mathbf{c}$ here denotes the indexes of the frames and the variables. 
Here, $\mathbf{x}^{\mathrm{c}}$ denotes the given history frames (e.g., the history monitoring results, including velocity and density changes, permeability, saturation, and RTM images) as well as the observed frames (e.g., current seismic observations) or variables (e.g., constant permeability). 
That means the number of frames and the number of variables used as conditions are not fixed, which will increase the amount of training data.

Thus, in real applications, given the history frames and measurements, and to achieve conditional sampling $\mathbf{x} \sim p_\theta\left(\mathbf{x} \mid \mathbf{x}^{\mathrm{c}}\right)$, we use reconstruction guidance \citep{ho_video_2022}.
As opposed to incorporating the conditions directly in the training process, we utilize a previously trained unconditional diffusion model to sample data given $\mathbf{x}_t$ at each diffusion step. We then estimate the residuals between the conditions $\mathbf{x}^\mathbf{c}$ and the sampled results at the indices of the history states and/or observations. This is done by minimizing the $l_2$ norm of the reconstruction loss:
\begin{equation}
\left\|\mathbf{x}^{c}-\widehat{\mathbf{x}}_\theta^c\left(\mathbf{x}_t, t\right)\right\|_2^2.
\end{equation}
This approach allows us to effectively leverage the unconditional model while aligning the generated data with the given conditions.

Then, we can calculate the gradient of this loss function with respect the $\mathbf{x}_b$, $\nabla_{\mathbf{x}_t^b}\left\|\mathbf{x}^{c}-\hat{\mathbf{x}}_\theta^c\left(\mathbf{x}_t, t\right)\right\|_2^2 $, to correct the sampling projection in each diffusion step, where the superscript $\mathbf{b}$ denotes the indexes of the target prediction complement to $\mathbf{c}$. 
Note that, unlike the usual way of calculating the gradient with respect to $\mathbf{x}_t^c$ or the whole sample $\mathbf{x}_t$ at time step t, instead, the gradient here is with respect to $\mathbf{x}_b$. 
The reason behind this formulation is that we hope to use the guidance, which represents the prediction at the conditional indexes $\mathbf{c}$ and should be consistent with the given conditions, to alter the sampling within the manifold of the unconditional region to the sampling within the manifold shifted by the guidance.
Our conditional diffusion model with reconstruction guidance is displayed in Figure~\ref{fig:diagram}.
Hence, the new sampling procedure of the reverse diffusion process is changed from the original $\hat{\mathbf{x}}_\theta\left(\mathbf{x}_t,t\right)=(\mathbf{x}_t - \sqrt{1-\bar{\alpha_t}}\widehat{\boldsymbol{\epsilon}}_{\boldsymbol{\theta}}\left(\mathbf{x}_t,t\right))/\sqrt{\bar{\alpha}_t}$ to 
\begin{equation}
\tilde{\mathbf{x}}_\theta^b\left(\mathbf{x}_t,t\right)\leftarrow\hat{\mathbf{x}}_\theta^b\left(\mathbf{x}_t,t\right)-\frac{w_r \sqrt{\bar{\alpha}_t}}{2} \nabla_{\mathbf{x}_t^b}\left\|\mathbf{x}^{c}-\hat{\mathbf{x}}_\theta^c\left(\mathbf{x}_t,t\right)\right\|_2^2,
\end{equation}
where the additional term is the correction from the reconstruction loss, $w_r$ is the weighting factor to control the strength of the guidance, and $\sqrt{\bar{\alpha}}_t$ is to control the guidance at different reverse diffusion steps.
The predictions at the conditional indexes $\mathbf{c}$ are replaced by the conditions.
With this framework, for example, if the network is trained on a small subset of six frames for computational efficiency, given the three history frames $\mathbf{x}_c$, we can predict the next three frames. 

Besides the feasibility of the forecasting and inversion, with the reconstruction guidance, we can theoretically extend the multiphysics evaluation to an arbitrary length by autoregressive conditional generation. 
That means we can take the prediction $\mathbf{x}^{\frac{T}{2}:T}$ as the history states and use them as the condition to forecast the $\mathbf{x}^{T:\frac{3T}{2}}$. Then, by iteratively achieving this prediction, we can extend the forecasting to the out-of-scope distributions.

\subsection{Uncertainty analysis}
As mentioned above, we utilize the diffusion model to capture the joint distribution of multiphysics evolution. 
Then, starting from a batch of random noise as priors, we can generate multiple samples/realizations guided by the same observations and history states.
Specifically, with the reconstruction guidance, we can generate samples that are all almost consistent with the conditions but slight variations may exist in areas not constrained well by the observations.
We can measure the mean and variance of the generated samples as an estimate of uncertainty.
Thus, in this paper, we will also showcase the estimated uncertainty using the proposed method in Section~\ref{sec:exp}.

\subsection{CO$_2$ related Multiphysics datasets generation} 
\begin{figure}[!htb]
    \centering
    \includegraphics[width=1.0\textwidth]{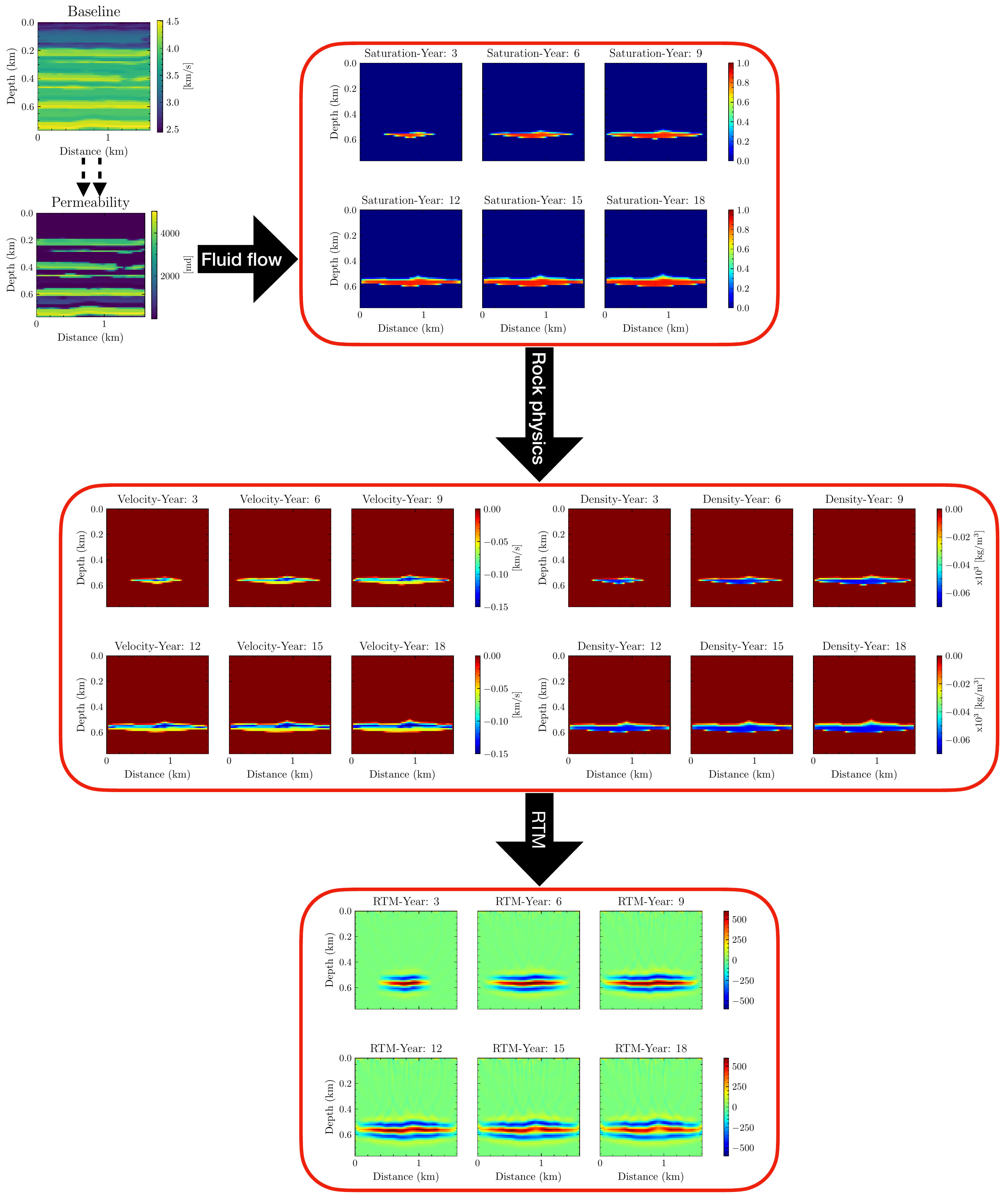}
    \caption{A multiphysics forward modeling framework starting with a permeability, which is converted from the baseline velocity model and then used to establish the CO$_2$ saturation given the constant injection using fluid flow simulation, which is in turn used to establish the changes in velocity and density based on rock physics modeling.  Finally, the RTM images are obtained from the simulated data using the evolved velocity and density models.}
    \label{fig:workflow}
\end{figure}

The training data, especially those that offer accurate evolution of the multiphysics properties during CO$_2$ injection, define the success of diffusion models, as they store the data distribution. However, in this study, our aim is to test the concept, so we make some simplifications and assumptions in our preparation of the training data.
Since our objective is to demonstrate and realize an end-to-end multiphysics inversion and prediction, we prepared our training datasets based on multiphysics forward modeling techniques, as detailed in \cite{louboutin_learned_2023}, and applied these techniques on synthetic data corresponding to the  Compass velocity models. The forward process encompasses several stages: fluid flow simulation, followed by rock physics modeling, seismic modeling, and finally, reverse time imaging.  
For more intricate details of each stage in the forward process, we refer readers to \cite{yin_slimgroupseis4ccsjl_2022,yin_derisking_2023}.
Additionally, Figure~\ref{fig:workflow} illustrates the workflow used to prepare the training datasets.

\textbf{Fluid flow simulation.} 
Before we can simulate fluid flow, we need to define the permeability $K$ of the subsurface model. Here, as suggested by \cite{yin_derisking_2023,louboutin_learned_2023}, we convert the wave speed $v$ of the compass model to log-permeability values to make up alternating high- and low-permeability layers in the reservoir with a seal on top. 
The conversion follows a non-linear relationship, where the permeability $K$ is defined as follows: $K=3000e^{(v-4)}$ for $v\geq4$; $K=0.01e^{(25.223(v-3.5))}$ for $3.5\leq v<4$; and $K=0.01e^{(v-3.5)}$ for $v<3.5$. 
This permeability is generated once based on velocity and kept constant regardless of the velocity changes due to CO$_2$ injection.
The porosity $\phi$ is also set to a constant value of 0.25.
Then, we place one injection well in the deeper part of the target domain and consider a constant injection rate for the experiment here. 
We use a two-phase simulation to obtain the CO$_2$ saturation model in Figure~\ref{fig:workflow} with softwares Jutul.jl \citep{moyner_sintefmathjutuljl_2023} and JutulDarcyRules.jl \citep{yin_slimgroupjutuldarcyrulesjl_2023}.

\textbf{Rock physics modeling.} 
Since we use seismic methods to monitor the subsurface CO$_2$ changes, we need to obtain the velocity changes resulting from injection and the changes in saturation.
We use the patchy saturation model \citep{louboutin_learned_2023} to convert the saturation model to velocity and density models, in which an increase in CO$_2$ saturation will result in a decrease in velocity and density. 
We ignore the pressure effect here and assume the shear wave velocity is fixed as $\frac{v}{\sqrt{3}}$, and the non-linear conversion follows
\begin{equation}
\begin{array}{ll}
\hat{\rho} & =\rho-0.276\phi S\\
\hat{v} & =\sqrt{{\frac{1}{{\hat{\rho}}}\left[(1-S)\left(\rho v^2\right)^{-1}+S\left(
\frac{B_{min}}{ (

\frac{\rho v^2}{1.8B_{min}-\rho v^2}-

\frac{2.735}{\phi\left(B_{min}-2.735\right)}+

\frac{0.125}{\phi\left(B_{min}-0.125\right)}

)^{-1}+ 1}
+\frac{4}{9} \rho v^2\right)^{-1}\right]^{-1}}}
\end{array}
    \label{equ:patchy}
\end{equation}
where $S$ is saturation, $v$ and$\rho$  are the velocity and density of the baseline model, $\hat{v}$ and $\hat{\rho}$ are the changed velocity and density after CO$_2$ injection, respectively, and $B_{min}=50$ for $v\geq3.5$ and  $B_{min}=\frac{2}{3}\rho v^2$ for $v<3.5$ km/s.

\textbf{Seismic modeling and reverse time imaging.} 
After obtaining the seismic properties changes represented by the velocity and density, we simulated the seismic recordings based on the impedance model calculated from the velocity and density, and through using born modeling. We then applied RTM to obtain seismic imaging results  \citep{yin_derisking_2023} by the software JUDI.jl \citep{louboutin_slimgroupjudijl_2023}. 
The details on the source and receiver locations will be shared in the examples section. 
Here, we use RTM images as the seismic response for convenience in testing our concept.
As the seismic responses to the velocity and density changes are quite small compared to the baseline monitoring, we do not use the RTM images directly. Instead, we use the residuals between the current RTM image and the baseline RTM image as the seismic observation. The Born modeling here will secure for us clean images as if processing and demultiple were applied to conventionally observed data.

\subsection{The backbone network and training configuration}
The backbone of the video diffusion model can have any architecture, e.g., 3D-Unet based \citep{cicek_3d_2016} or transformer based \citep{peebles_scalable_2022}. 
In this paper, since we do not use a variational autoencoder to compress the videos to the latent space, using a transformer would be costly due to the quadratic computational complexity in the attention mechanism of the transformer.
Thus, we use a 3D-Unet, which is factorized over space and time. Compared to the 2D Unet often used for image processing, whose convolutional layer is often with the kernel size of 3$\times$3, we use 1$\times$3$\times$3 convolutional kernel for the video. 
The dimensions of the kernel correspond to the time frames, height, and width of the multiphysics evolution.
The attention in spatial attention block used in the 3D-Unet applies to the space dimension only, in which the time frame axis is regarded as the batch dimension, combined with a following temporal attention block. 
To distinguish the order of the frames (the time order of the multiphysics evolution), the relative position embeddings \citep{shaw_self-attention_2018} are used in each temporal attention block.
This paper's backbone network of the video diffusion model is designed to consider five channels, including permeability, CO$_2$ saturation, velocity, density, and RTM images. 

The number of the timesteps $T$ for the video diffusion is 1000, and the loss function of the training is given by the $l_1$ norm. 
To save on memory and stabilize the training, we use a gradient accumulation of 2, as well as mixed precision techniques \citep{micikevicius_mixed_2018}.
As has been used in many diffusion model training, we apply the Exponential Moving Average (EMA) \citep{xu_towards_2024} to smooth the parameters update of the neural network, improve the generalization, and accelerate the convergence.

\section{Numerical Experiments}
\label{sec:exp}
\begin{figure}
    \centering
    \includegraphics[width=0.9\textwidth]{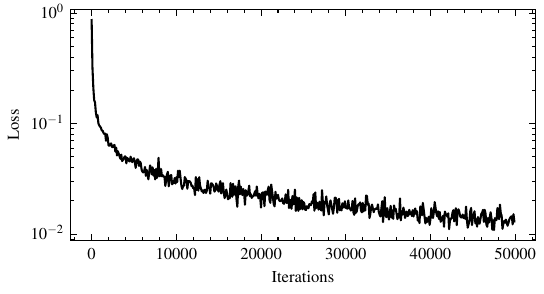}
    \caption{The training loss curve of the video diffusion model for multiphysics evolution.}
    \label{fig:loss_curve}
\end{figure}

In our experiments, we employed the 3D Compass velocity model, a synthetic model that closely mimics the geological characteristics found in the southeastern region of the North Sea. This model serves as an ideal testbed due to its realistic representation of the subsurface geological features.
Since we focus more on the demonstration of the framework, we extract 514 two-dimensional slices from the central deep part of the three-dimensional compass velocity model, having dimensions of $64\times64$, with an interval of 25 $m$ in the horizontal direction and 12 $m$ in the vertical direction. 
In this paper, we do not consider the leakage scenario and assume the permeability is constant over time, though this scenario can also be incorporated if needed. 
For the fluid flow simulation, we control the CO$_2$ injection rate to achieve a 10\% storage capacity over an 18-year period. 
The constant CO$_2$ injection happens at the same fixed well location for all cases (often known in field experiments).
From this simulation, we sampled six frames starting from the third year with a time-lapse interval of three years. 
These frames were crucial in analyzing the temporal evolution of the injected CO$_2$. 
Subsequently, we applied rock physics modeling to calculate the corresponding changes in velocities and densities due to CO$_2$ injection for each CO$_2$ saturation at different times.
We collected the simulated seismic data for each velocity and density model by uniformly placing six sources on the surface with a 24 Hz Ricker wavelet and 200 receivers on the sea bottom. 
The recordings lasted for 1 $s$ with a sampling rate of 4 $ms$.
Then, we employ RTM to compute the time-lapse seismic image changes by subtracting them from the baseline image. 
In the video diffusion training phase, which is conducted on two Nvidia A100 GPUs for 32.8 hours, we used an Adam optimizer with a learning rate of 5e-5, a batch size of 128, and a maximum of 50,000 iterations until the neural network converged, as evidenced by the training loss curve, shown in Figure~\ref{fig:loss_curve}.

\subsection{Unconditional monitoring sampling} 
\begin{figure}[!htb]
    \centering
    \includegraphics[width=0.76\textwidth]{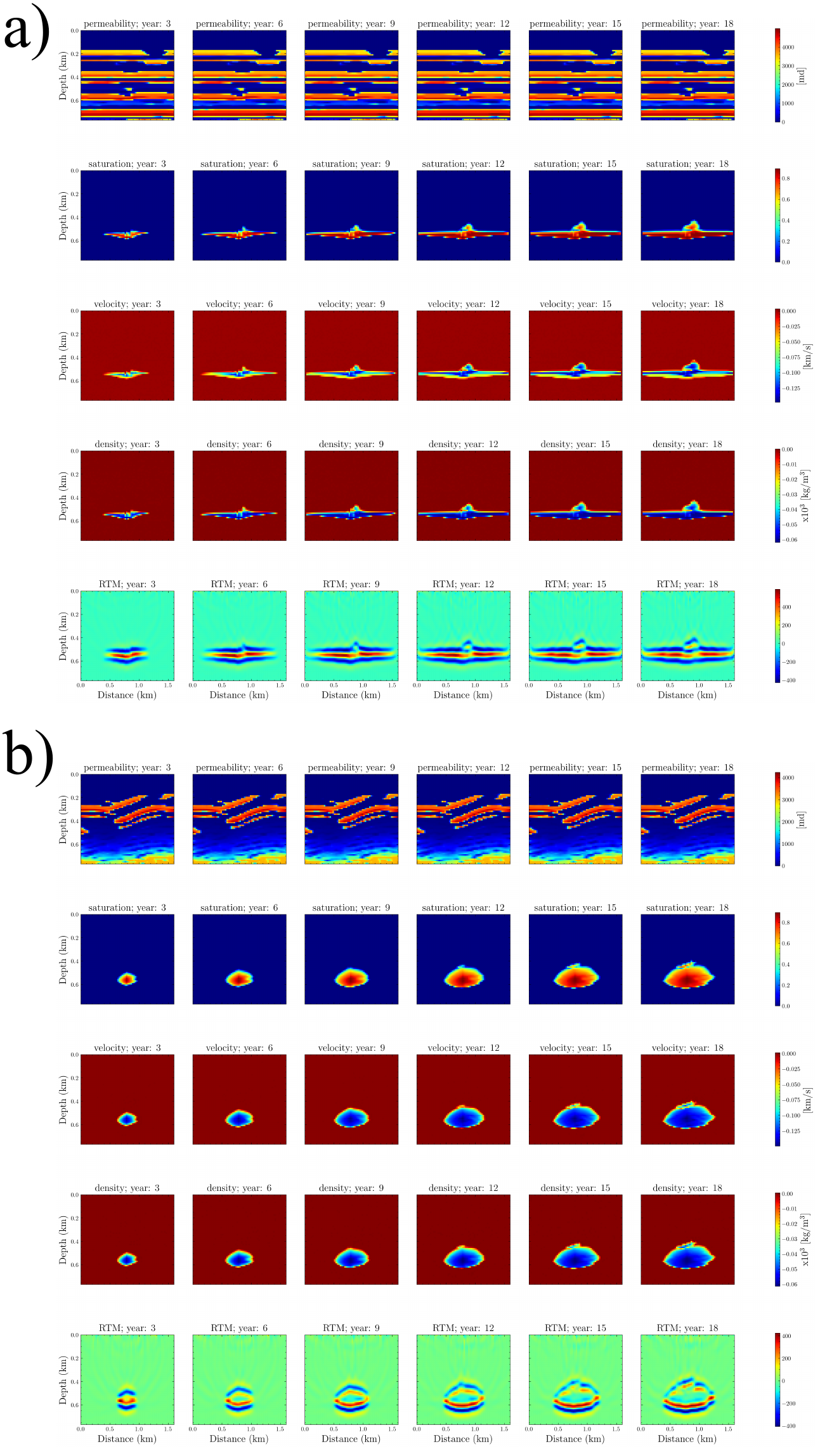}
    \caption{Uncodintional generation of multiphysics evaluation due to the CO$_2$ injection for two examples (a and b). Different frames denote different times with an interval of 3 years. Each row denotes a specific variable, including permeability, CO$_2$ saturation, velocity, density, and RTM images from top to bottom, respectively, while each column denotes the variables at different times.}
    \label{fig:uncoditional}
\end{figure}

After completing the training of our diffusion model, we conducted an initial test by sampling multiphysics evolution over time without any conditions. 
Inputting two random noise samples in the latent space of the Diffusion model, we use equation~\ref{equ:sample} to do the reverse sampling for 1000 time steps.
The results of this process are illustrated in Figure~\ref{fig:uncoditional}. As expected, we can see that starting from random Gaussian noise, we are able to generate high-quality spatial-temporal series for multiphysics evolution. The series showcased clear and detailed progression and maintained consistency with the underlying physical laws. 
It demonstrates that the diffusion model has successfully captured the inherently complex physical phenomena associated with CO$_2$ monitoring. 

\subsection{Forecasting} 
\label{sec:exp-forecasting}
\begin{figure}[!htb]
    \centering
    \includegraphics[width=0.76\textwidth]{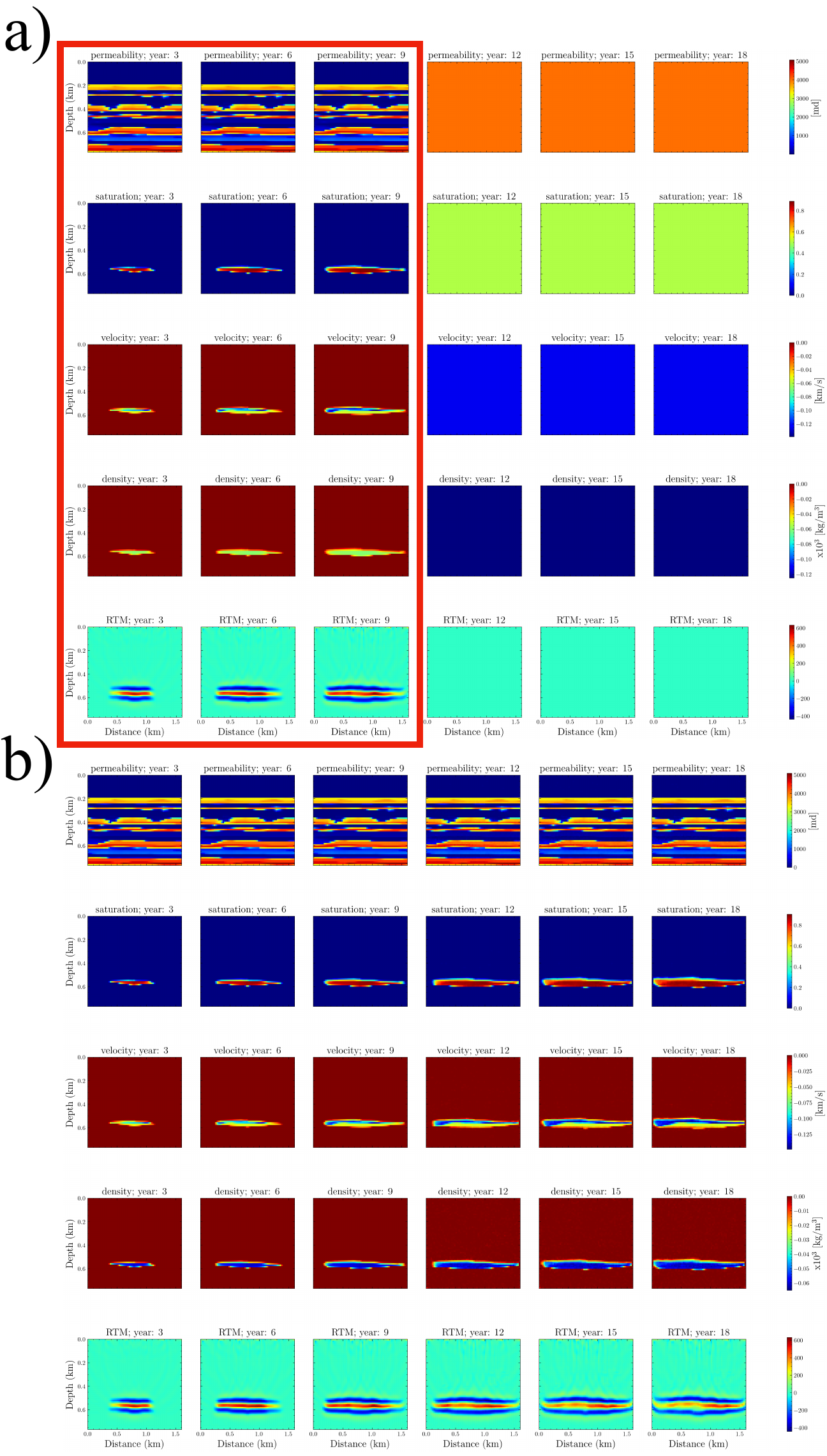}
    \caption{Forecasting generation (b) given three history frames (a), which are denoted by the red box. 
    The solid color subfigures in (a) denote the unknown future states, which are the target of the forecasting generation.
    The rows represent properties described in Figure \ref{fig:uncoditional}.}
    \label{fig:forecasting}
\end{figure}

\begin{figure}[!htb]
    \centering
    \includegraphics[width=0.76\textwidth]{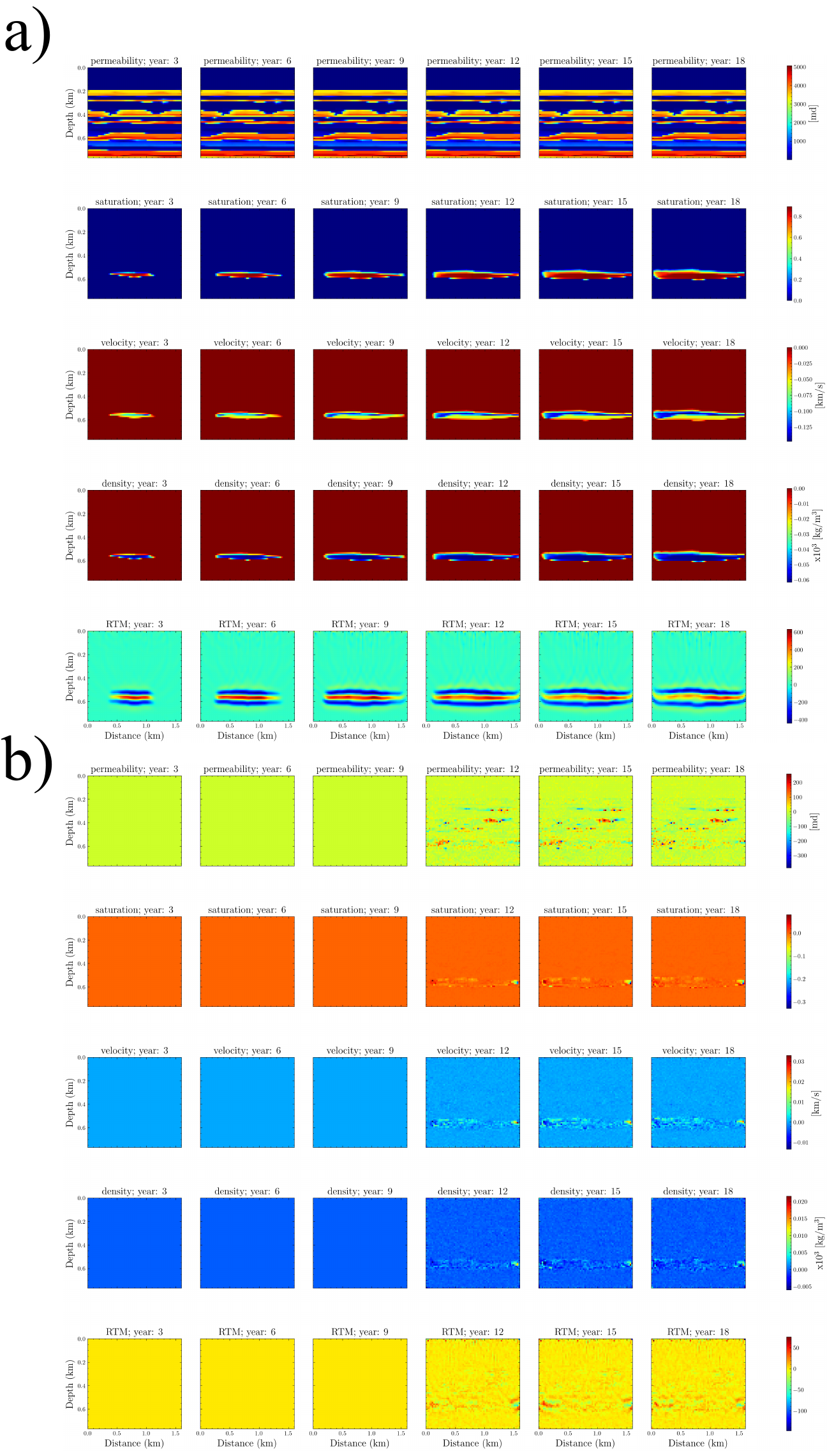}
    \caption{The errors (b) in forecasting generation (Figure~\ref{fig:forecasting}b) compared to the reference (numerical) result (a). The rows represent properties described in Figure \ref{fig:uncoditional}. The errors (differences) are plotted at the same scale.}
    \label{fig:forecasting_res}
\end{figure}

As mentioned in Section~\ref{sec:method}, this well-trained unconditional video diffusion model can be used for forecasting and inversion. 
We first test it for the forecasting task to evaluate the ability to predict future CO$_2$ related multiphysics evolution based on the history frames as that can help us evaluate the capacity for CO$_2$ storage for decision-making optimization. 
With a reconstruction scale $w_r$ of $10^6$, the conditional sampling results are shown in Figure~\ref{fig:forecasting}. 
Given the observations and records of the subsurface states in years 3, 6, 9, we forecast the future states for years 12, 15, and 18. 
They are consistent and reasonable in their evolution and generally maintain the coupling relationship between the five physical parameters.
To quantitate the forecasting errors for each variable, we compare them with the reference numerical simulations, shown in Figure~\ref{fig:forecasting_res}. 
The errors are small and are focused mainly in the region of the CO$_2$ plume. 
This generally validates our model's accuracy and highlights its potential as a powerful tool for strategic planning in CO$_2$ storage projects.
We test the forecasting performance on five distinct examples, which are not used during the training. 
The average relative L2 norm error, which is calculated by dividing the L2 norm of the error by the L2 norm of the ground truth and the lower value denotes a better performance, is 0.024, and the average structural similarity index measure (SSIM, the higher value denotes a better performance) \citep{zeng_3d-ssim_2012} is 0.889.  
\begin{figure}[!htb]
    \centering
    \includegraphics[width=0.76\textwidth]{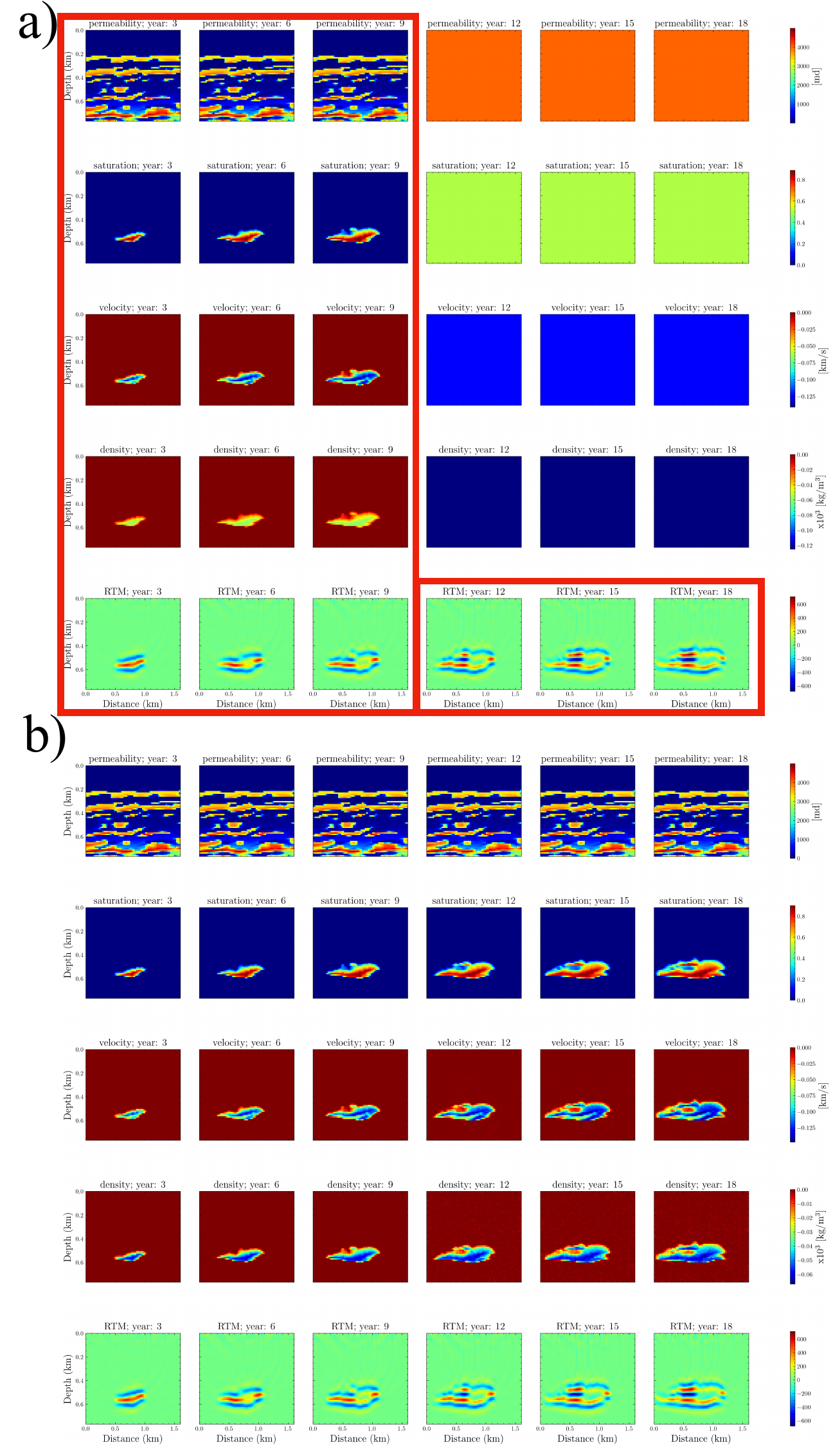}
    \caption{Inversion results (b) given three history frames and observed-data-based RTM images (a), which are denoted by the red box. 
    The solid color subfigures in (a) denote the unknown partial physical variables, which are the target of the inversion.
    The rows represent properties described in Figure \ref{fig:uncoditional}}
    \label{fig:inversion}
\end{figure}
\begin{figure}[!htb]
    \centering
    \includegraphics[width=0.76\textwidth]{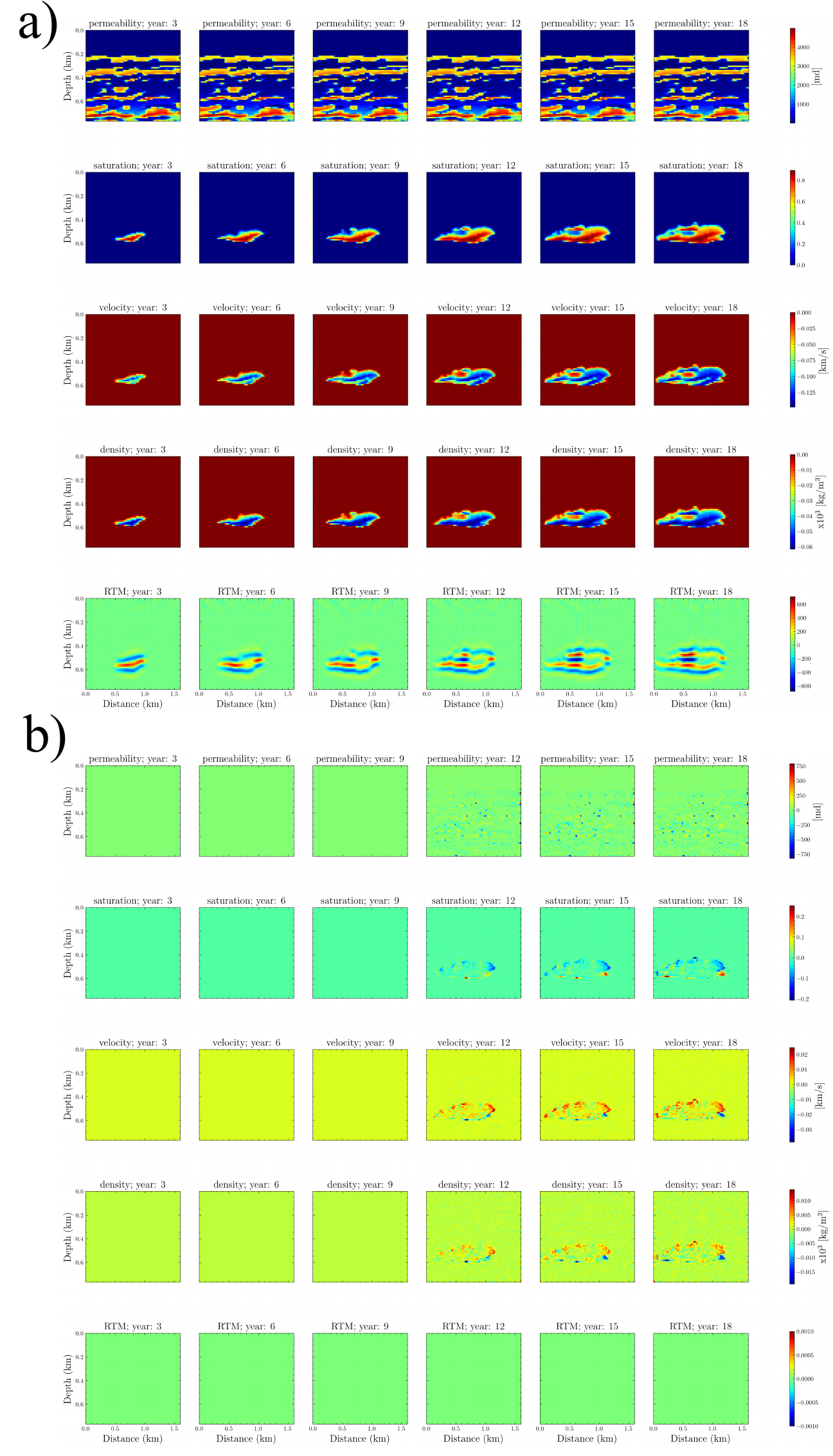}
    \caption{The errors (b) in the inverted results (Figure~\ref{fig:inversion}b) compared to the reference result (a). The rows represent properties described in Figure \ref{fig:uncoditional}. The errors (differences) are plotted at the same scale.}
    \label{fig:inversion_res}
\end{figure}

\subsection{Inversion} 
\label{sec:exp-inversion}
In realistic monitoring scenarios, besides the historical information of the multiphysics evolution, we might also have some measurements of the seismic response on the surface, such as the seismic recordings, which can be converted to RTM images \citep{li_neural_2021}. 
RTM images play a crucial role in these instances, acting as a strong conditioner that can constrain and guide the generation of CO$_2$ evolution and the corresponding changes in subsurface elastic properties.
With our method, we can do an inversion of the subsurface velocity, density, and CO$_2$ saturation, given the observed seismic images. 
To evaluate the effectiveness of this approach on inversion, we conducted tests where the generation process was conditioned by both historical frames and observed RTM image results. 
The inverted results are shown in Figure~\ref{fig:inversion}. 
We observed that the diffusion models can invert high-quality subsurface parameters, and the evolution of the inverted velocity, density, and CO$_2$ saturation are reasonable and consistent.
We also quantify the errors of the inverted results using the reference numerical simulations, shown in Figure~\ref{fig:inversion_res}. 
Compared to the reference results, our method showcases a good inversion performance in terms of accuracy, especially for the velocity and density models, which the images are more dependent on. 
This makes sense as the more conditions we have, like the RTM image results, which can provide strong constraints on velocity and density, the better the accuracy of the prediction of our method.
We also test the inversion performance on five distinct examples. 
The average relative L2 norm error is 0.021, and the average SSIM is 0.961. 

\begin{figure}[!htb]
    \centering
    \includegraphics[width=0.76\textwidth]{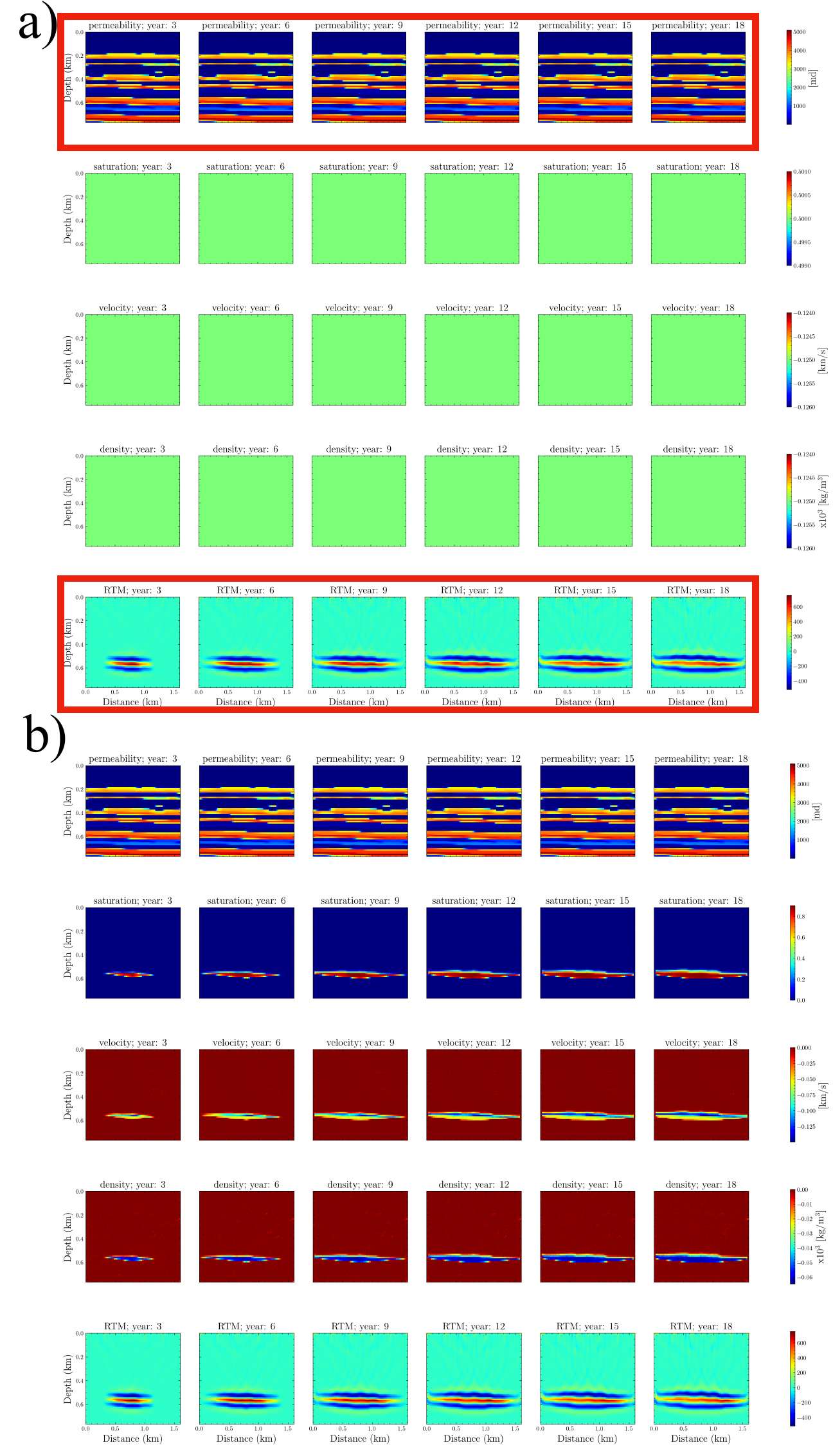}
    \caption{Inversion results (b) given permeability and observed-data-based RTM images (a), which are denoted by the red box. 
    The solid color subfigures in (a) denote the unknown partial physical variables, which are the target of the inversion.
    The rows represent the various monitored properties described in Figure \ref{fig:uncoditional}}
    \label{fig:inversion_2}
\end{figure}

\begin{figure}[!htb]
    \centering
    \includegraphics[width=0.76\textwidth]{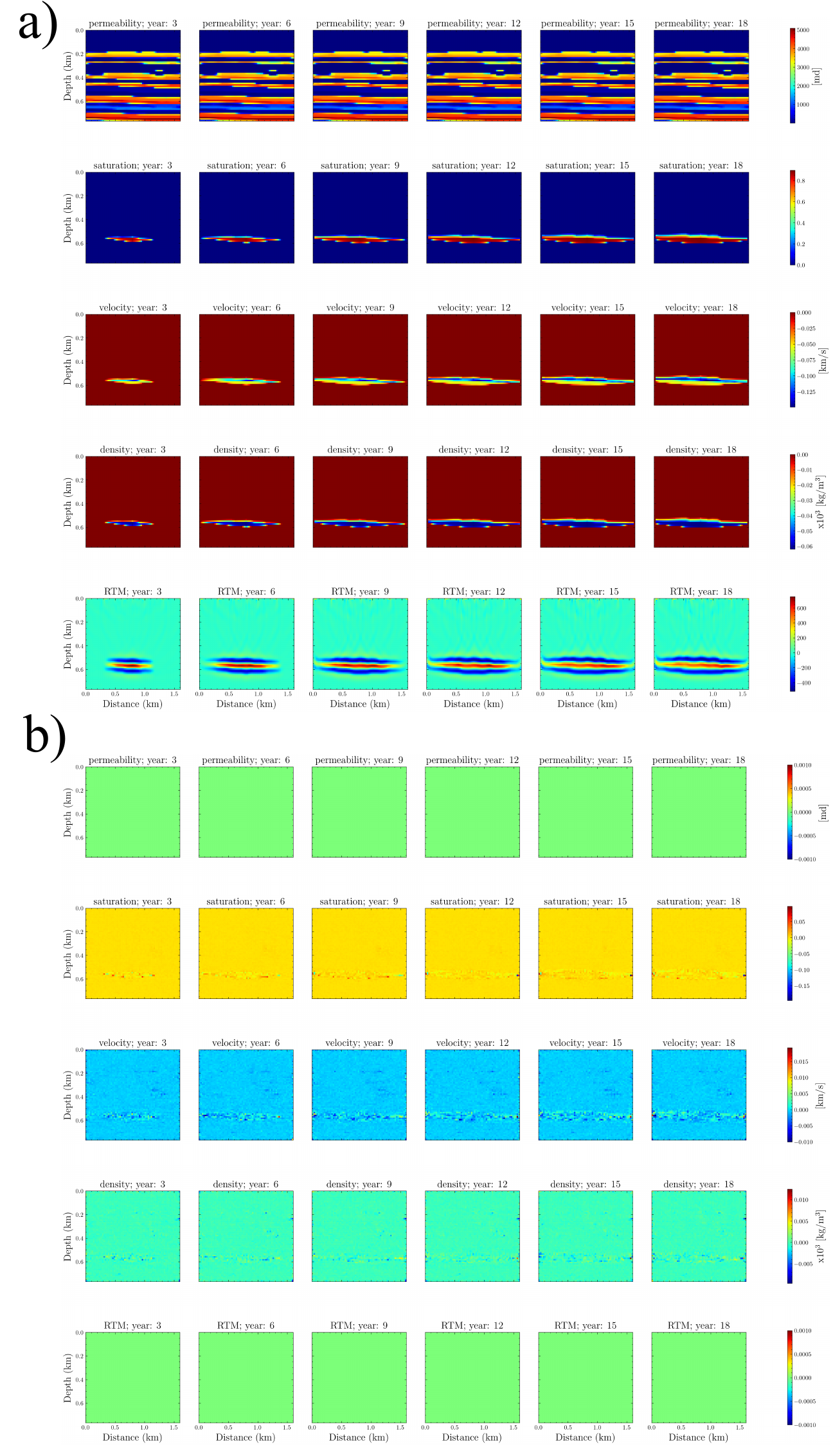}
    \caption{The errors (b) in the inverted results (Figure~\ref{fig:inversion_2}b) compared to the reference result (a). The rows represent the various monitored properties described in Figure \ref{fig:uncoditional}. The errors (differences) are plotted at the same scale.}
    \label{fig:inversion_2_res}
\end{figure}

Another scenario of inversion is that we only have the information on permeability and observations like RTM images and try to estimate the CO$_2$ saturation and elastic parameters. The inverted results using our proposed method are shown in Figure~\ref{fig:inversion_2}, and the quantitative analysis of the errors is shown in Figure~\ref{fig:inversion_2_res}. 
They show good reconstruction of the velocity, density, and CO$_2$ saturation evolutions.
In such a scenario, the average relative L2 norm error of five distinct examples is 0.083, and the average SSIM is 0.984.

\subsection{Out-of-scope forecasting}
\begin{figure}[!htb]
    \centering
    \includegraphics[width=0.76\textwidth]{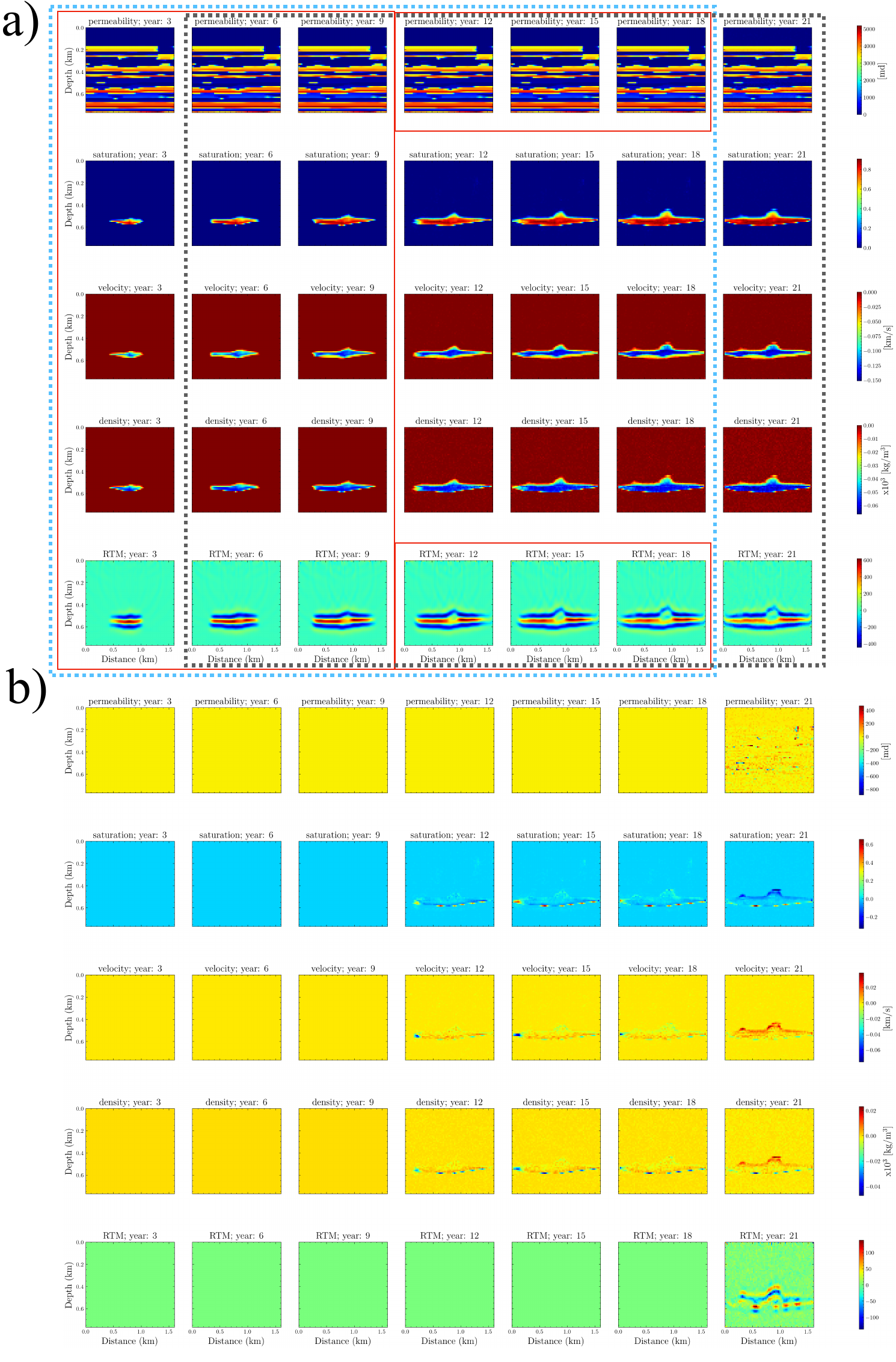}
    \caption{Out-of-horizon autoregressive predictions (a) given three history frames at 3, 6 and 9 years, and permeability and RTM images at 3 to 18 years, which are denoted by the red box. The first prediction is denoted by the blue dashed box while the second autoregressive prediction is highlighted by the gray dashed box. (b) shows the errors between the forecasting results and numerical simulated ones.}
    \label{fig:ood-forecasting}
\end{figure}

Combining the unconditional video diffusion model and reconstruction guidance, our framework can forecast not only the temporal scope of in-distribution but also the forecasting of the out-of-distribution scope by autoregressive conditional sampling.
In theory, the proposed method can expand the evolution to arbitrary length. 
However, due to the limits of the spatial size of our simulation domain, the CO$_2$ plume at year 24 may exceed the boundary of the simulation domain, which is not the case in realistic applications because our monitoring domain is usually bigger than the CO$_2$ plume.
Hence, we first use the frames from 3, 6, and 9 years as history frames and the observations like RTM images and permeability (denoted by the red box in Figure~\ref{fig:ood-forecasting}a) to invert the in-distribution scope (denoted by the blue dashed box in Figure~\ref{fig:ood-forecasting}a) and, then use the states at 6, 9, 12, 15, and 18 years as history states to predict the future states at 21 years (denoted by the gray dashed box in Figure~\ref{fig:ood-forecasting}a) for a proof-of-concept test. 
In addition, to also test the generalization ability, the injection rate for this sequence is decreased to achieve a 10\% storage capacity over a 21-year period.
As shown in Figure~\ref{fig:ood-forecasting}a, the diffusion model captures not only the dynamics over different physical variables but also the dynamics of evolution over time. The predictions over the horizon of training are still reasonable, demonstrating the potential of the proposed method for forecasting beyond the trained horizon. 
Then, we also simulated the reference results for the comparison. The errors shown in Figure~\ref{fig:ood-forecasting}b highlight the ability of the proposed framework in the out-of-distribution forecasting task.
To improve forecasting accuracy for longer sequences or different injection rates, one can incorporate more time frames during the training while keeping the number of frames for each sample and more sequences using different injection rates.

\subsection{Uncertainty analysis}
\begin{figure}[!htb]
    \centering
    \includegraphics[width=0.76\textwidth]{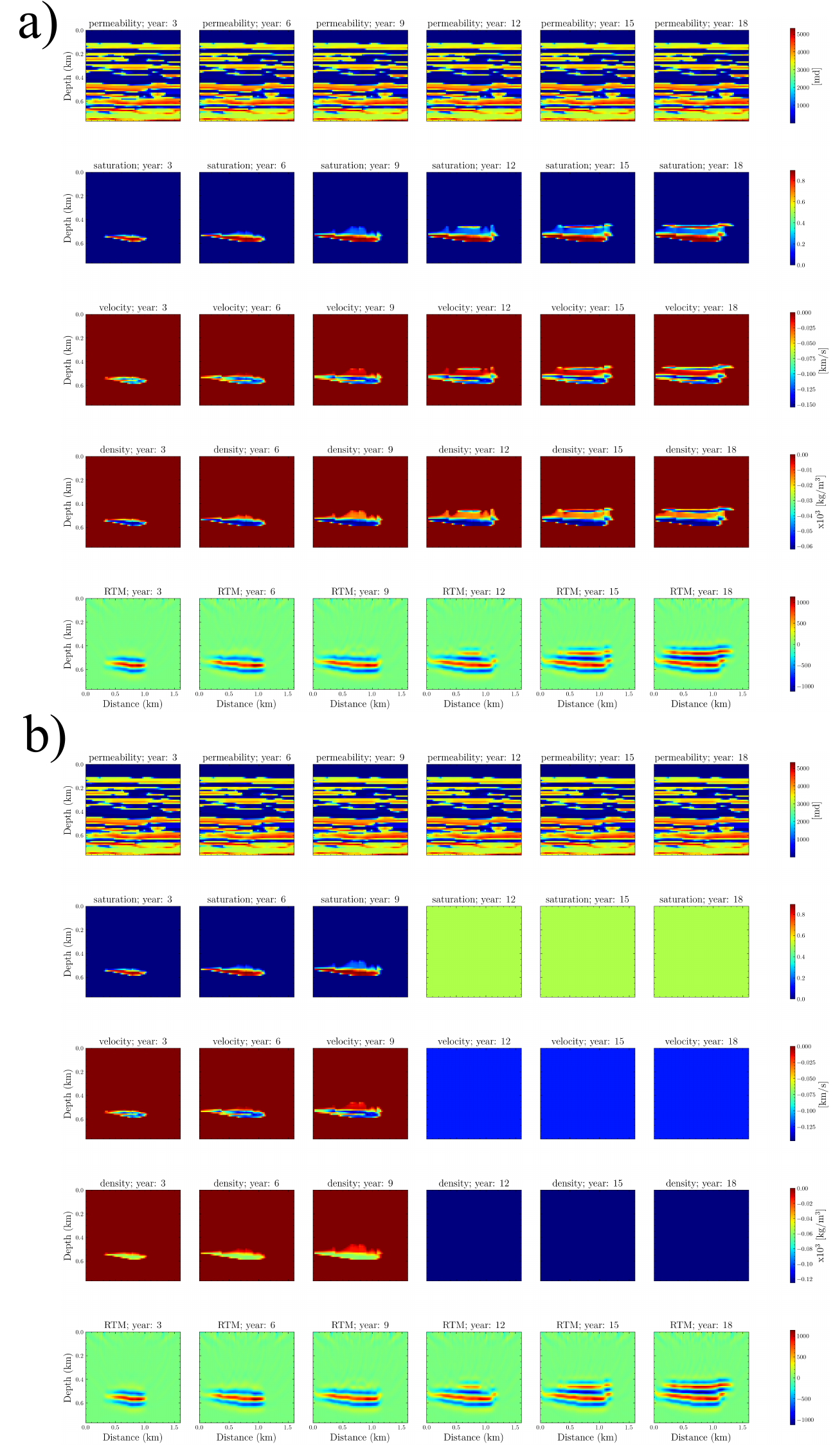}
    \caption{The reference numerical simulated results (a) and the given conditions used for the conditional generation (b). The solid color subfigures in (b) denote the unknown partial physical variables, which are the target of the inversion.}
    \label{fig:uncertainty_1}
\end{figure}

\begin{figure}[!htb]
    \centering
    \includegraphics[width=0.76\textwidth]{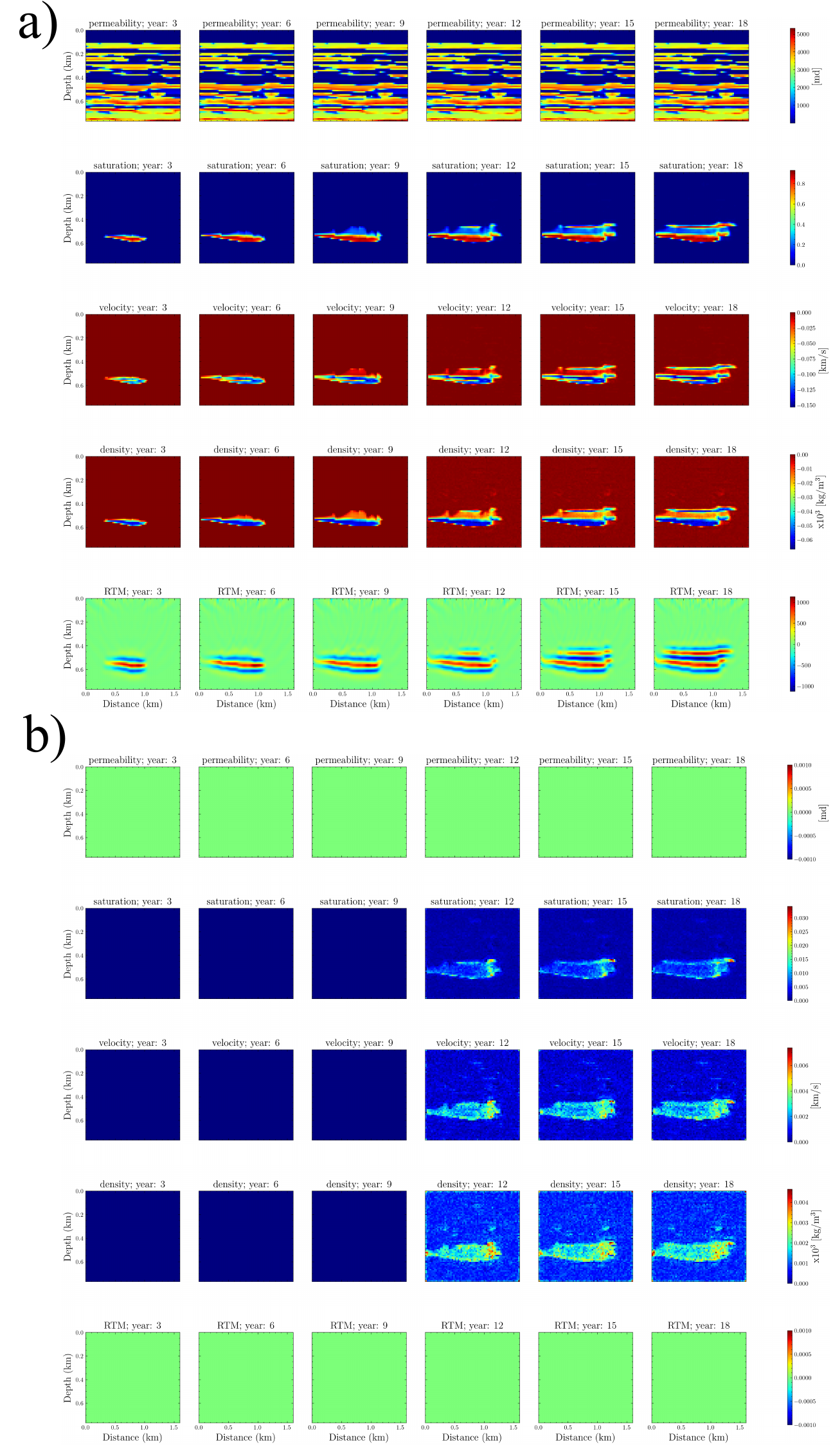}
    \caption{The mean prediction of 16 inverted results (a) given three history frames, constant permeability, and observed RTM images starting from different random noises. (b) shows the standard deviation of these 16 inverted results.}
    \label{fig:uncertainty_mean_std}
\end{figure}

Despite the generally high accuracy we managed to achieve in forecasting and inversion, we still had errors as compared to the numerical simulation the training of the diffusion models was based on. As we saw, the accuracy depends on the amount of data we use to condition the generation. Thus, uncertainty quantification as an additional output of the framework can help us assess confidence in the inversion and forecasting results in practical applications, where reference solutions are unavailable.
Recall that our monitoring framework is based on the video diffusion model, which is one type of generative models that naturally supports the uncertainty analysis \citep{finzi_user-defined_2023,price_gencast_2023}. 
The way to quantify approximate uncertainty within our framework is generally simple and cheap.
We mainly need perform conditional sampling for the same history states and measurements starting from several random noise samples in parallel, let's say 16 samples for the sake of the test though this number is theoretically small to properly sample the posterior distribution.
Thus, we evaluate the mean and standard deviation of the generated 16 evolution samples. 
We assume a constant permeability in this test and give the three past history states and the RTM image results over the years to perform the inversion.
We show the reference numerical simulated results and the conditions used for generation in Figure~\ref{fig:uncertainty_1}. 
The mean prediction is shown in Figure~\ref{fig:uncertainty_mean_std}a.
Similar to the performance shown in Sections~\ref{sec:exp-forecasting} and \ref{sec:exp-inversion}, the sample shows reasonable inversion results, and the consistency of evolution between different variables remains. 
The average relative error is 0.18, and the average SSIM is 0.9815.
The standard deviation values for those 16 samples are shown in Figure~\ref{fig:uncertainty_mean_std}b.
As expected, high uncertainty appears at the boundary of the plume, especially around the potential leakage area, and decreases from the boundary to the interior of the CO$_2$ plume. 
In addition, such uncertainty patterns might indicate regions where predictions are less reliable and thus require additional measurements, like well logs. 
Interpreting uncertainty in this context enables more informed decisions for taking preventive measures.
Overall, the experiment demonstrates that the proposed method can achieve probabilistic predictions in a fully parallel way and can provide a reasonable approximate uncertainty map.
Unlike ensemble approaches \citep{fort_deep_2020}, which require training of multiple models, our approach can do the probabilistic generation within one model and is more efficient. 


\section{Dicussions}
\label{sec:discuss}
This paper presents a novel framework utilizing video diffusion models for subsurface forecasting and monitoring. It demonstrates good performance in terms of physical consistency, visual quality, and accuracy. 
However, a notable limitation is the computational cost of the sampling process, which necessitates iterative inference (1000 iterations in our case).
Generating one sample on a Nvidia A100 GPU will cost 29 seconds for unconditional generation and 84 seconds for conditional generation.
This issue highlights the need for further optimization to reduce the computational cost, possibly through approaches like denoising diffusion implicit models \citep{song_denoising_2022} or consistency models \citep{song_consistency_2023}. 
In addition, to handle the high-resolution multiphysics images over longer time-lapse evolution, where the computational cost is expected to be huge, and the learning complexity increases, advanced techniques such as stable diffusion \citep{rombach_high-resolution_2022} and \citep{peebles_scalable_2022} are needed. 

Besides the computational cost, to apply our method in field applications, addressing the gap between the training data and realistic multiphysics evolution is required. 
Specifically, the permeability in this paper is assumed to be constant for simplification.
Although incorporating this constant channel as additional information for different CO$_2$ related multiphysics evolution has shown good performance in this paper, it might also have time-dependent properties in realistic scenarios. 
Hence, we can consider the time-dependent permeability in the training data preparation to make it better resemble a realistic scenario.
The heterogeneous geological formations should also be considered during the data generation for complex field applications.
Furthermore, we can add noise into the seismic recordings and adjust the seismic acquisition before obtaining the corresponding RTM images to imitate noisy real-world observations. 
In addition, we may have a time-dependent injection rate rather than a constant one, as used in this paper. 
In some cases, we might stop the injection for a while and continue later.
What's more, the location of the injection well and multiple injection wells should also be considered.
Thus, including such a variable would be more interesting and important. 
The classifier-free conditional diffusion model \citep{ho_classier-free_2021,wang_controllable_2024} would be a potential solution, which we might explore more in the future. For example, we can have the injection rate over a period as a condition, as such information is often known.

In the current pipeline, we use RTM images as the seismic response rather than the seismic recordings. The adaption to seismic recordings also requires video diffusion models to learn a mapping from the velocity and density to the seismic wavefield. This mapping, as we know from full waveform inversion, is very complicated, requiring an abundance of training data, and often lacks generalization abilities.
Although the ideal workflow is to enable the inversion or forecasting based on the observed seismic data, which is quite straightforward in realistic monitoring scenarios, such an undertaking will increase the complexity of the whole multiphysics process further as we need to teach the neural network to learn the wave-equation modeling engine. 
This challenge reveals another interesting future direction, in which we can try to learn a neural PDE solver \citep{huang_lordnet_2024,huang_learned_2024} for instant seismic modeling given the velocity and geometry or directly incorporate the differential seismic wavefield modeling engine, such as DeepWave \citep{richardson_deepwave_2023}, into the monitoring framework.

Unlike neural PDE solvers, which can learn simulation mappings between the given input multiple variables and the corresponding CO$_2$ saturations, velocity, and so on, the proposed framework directly learns the end-to-end multiphysics evolution by processing them as video.
Since the physical world and complex physical phenomena like CO$_2$ multiphysics evolution can be hypothetically filmed and captured as a video, using a video diffusion model to store such evolution, including the physics involved, seems natural.
Beyond the demonstration in this paper for this concept, there are also many literatures showing that the video diffusion models can simulate the world \citep{cachay_dyffusion_2023,price_gencast_2023,ding_diffusion_2024}. However, to make the diffusion models store the real physics and generate the evolution with arbitrary length, a large amount of training samples are needed. 
In addition, as shown in the paper, the forecasting results still exhibit errors compared to the reference results. 
To further enhance the performance and reduce the number of training samples, we believe that incorporating the physical knowledge, for example, the governing equations as a physics-constraint loss \citep{huang_lordnet_2024,christopher_projected_2024}, in the sampling process can enhance the dynamic consistency over time, and correct the errors.
To improve the forecasting ability to longer future horizons, combining the next-token prediction models, which enable variable-length generation, with the video diffusion models, e.g., diffusion forcing \citep{chen_diffusion_2024}, might be a potential solution path for the framework.

\section{Conclusions}
\label{sec:conclusion}
We proposed a novel subsurface multiphysics monitoring and forecasting framework utilizing video diffusion models. Our method demonstrates the ability to generate high-quality representations of CO$_2$ evolution and associated changes in subsurface elastic properties. With reconstruction guidance, it shows promising results in forecasting and inversion based on historical frames and observational data. Additionally, thanks to the diffusion model’s inherent support for uncertainty quantification, our framework provides an efficient way for conducting reasonable uncertainty quantification.
The test on the Compass model demonstrates the effectiveness of the proposed method and the potential of the proposed method as the unified framework for forecasting, inversion, and uncertainty analysis of CO$_2$ related multiphysics monitoring and inversion.

\section{Open Research Section}
The datasets used in this paper can be downloaded at \cite{Huang2024diffusion-dataset}. During the dataset generation, we use Jutul.jl \cite[]{moyner_sintefmathjutuljl_2023} and JutulDarcyRules.jl \cite[]{yin_slimgroupjutuldarcyrulesjl_2023} for flow simulation and JUDI \cite[]{louboutin_slimgroupjudijl_2023} for RTM images generation.  
Codes of the video diffusion models to reproduce the results presented here are also available at \cite{Huang2024diffusion-dataset}.
Figures were made with Matplotlib version 3.2.1 \cite[]{caswell_matplotlibmatplotlib_2020}.

\section{Acknowledgments}
We thank KAUST and the sponsors of the Deepwave Consortium for their support.
We also thank Ziyi Yin for the discussion on data generation.  
This work utilized the resources of the Supercomputing Laboratory at KAUST, and we are grateful for that.
\bibliography{references}

\begin{thebibliography}{63}
\providecommand{\natexlab}[1]{#1}
\providecommand{\url}[1]{\texttt{#1}}
\expandafter\ifx\csname urlstyle\endcsname\relax
  \providecommand{\doi}[1]{doi: #1}\else
  \providecommand{\doi}{doi: \begingroup \urlstyle{rm}\Url}\fi

\bibitem[Blunt et~al.(1992)Blunt, King, and Scher]{blunt_simulation_1992}
M.~Blunt, M.~J. King, and H.~Scher.
\newblock Simulation and theory of two-phase flow in porous media.
\newblock \emph{Physical Review A}, 46\penalty0 (12):\penalty0 7680--7699, Dec. 1992.
\newblock \doi{10.1103/PhysRevA.46.7680}.
\newblock URL \url{https://link.aps.org/doi/10.1103/PhysRevA.46.7680}.
\newblock Publisher: American Physical Society.

\bibitem[Bosch et~al.(2010)Bosch, Mukerji, and Gonzalez]{bosch_seismic_2010}
M.~Bosch, T.~Mukerji, and E.~F. Gonzalez.
\newblock Seismic inversion for reservoir properties combining statistical rock physics and geostatistics: {A} review.
\newblock \emph{GEOPHYSICS}, 75\penalty0 (5):\penalty0 75A165--75A176, Sept. 2010.
\newblock ISSN 0016-8033.
\newblock \doi{10.1190/1.3478209}.
\newblock URL \url{https://library.seg.org/doi/10.1190/1.3478209}.
\newblock Publisher: Society of Exploration Geophysicists.

\bibitem[Cachay et~al.(2023)Cachay, Zhao, Joren, and Yu]{cachay_dyffusion_2023}
S.~R. Cachay, B.~Zhao, H.~Joren, and R.~Yu.
\newblock {DYffusion}: {A} {Dynamics}-informed {Diffusion} {Model} for {Spatiotemporal} {Forecasting}, Oct. 2023.
\newblock URL \url{http://arxiv.org/abs/2306.01984}.
\newblock arXiv:2306.01984 [cs, stat].

\bibitem[Caswell et~al.(2020)Caswell, Droettboom, Lee, Hunter, Firing, Stansby, Klymak, Hoffmann, Andrade, Varoquaux, Nielsen, Root, Elson, May, Dale, Lee, Seppänen, McDougall, Straw, Hobson, Gohlke, Yu, Ma, Vincent, Silvester, Moad, Kniazev, Ivanov, Ernest, and Katins]{caswell_matplotlibmatplotlib_2020}
T.~A. Caswell, M.~Droettboom, A.~Lee, J.~Hunter, E.~Firing, D.~Stansby, J.~Klymak, T.~Hoffmann, E.~S.~d. Andrade, N.~Varoquaux, J.~H. Nielsen, B.~Root, P.~Elson, R.~May, D.~Dale, J.-J. Lee, J.~K. Seppänen, D.~McDougall, A.~Straw, P.~Hobson, C.~Gohlke, T.~S. Yu, E.~Ma, A.~F. Vincent, S.~Silvester, C.~Moad, N.~Kniazev, P.~Ivanov, E.~Ernest, and J.~Katins.
\newblock matplotlib/matplotlib: {REL}: v3.2.1.
\newblock [Software], Mar. 2020.
\newblock URL \url{https://zenodo.org/records/3714460}.

\bibitem[Chen et~al.(2024{\natexlab{a}})Chen, Monso, Du, Simchowitz, Tedrake, and Sitzmann]{chen_diffusion_2024}
B.~Chen, D.~M. Monso, Y.~Du, M.~Simchowitz, R.~Tedrake, and V.~Sitzmann.
\newblock Diffusion {Forcing}: {Next}-token {Prediction} {Meets} {Full}-{Sequence} {Diffusion}, July 2024{\natexlab{a}}.
\newblock URL \url{http://arxiv.org/abs/2407.01392}.
\newblock arXiv:2407.01392 [cs].

\bibitem[Chen et~al.(2024{\natexlab{b}})Chen, Liu, Di, and Zhang]{chen_co2seg_2024}
G.~Chen, Y.~Liu, X.~Di, and H.~Zhang.
\newblock {CO2Seg}: {Automatic} {CO2} {Segmentation} {From} 4-{D} {Seismic} {Image} {Using} {Convolutional} {Vision} {Transformer}.
\newblock \emph{IEEE Transactions on Geoscience and Remote Sensing}, 62:\penalty0 1--14, 2024{\natexlab{b}}.
\newblock ISSN 1558-0644.
\newblock \doi{10.1109/TGRS.2024.3389780}.
\newblock URL \url{https://ieeexplore.ieee.org/document/10500845/ authors#authors}.
\newblock Conference Name: IEEE Transactions on Geoscience and Remote Sensing.

\bibitem[Christopher et~al.(2024)Christopher, Baek, and Fioretto]{christopher_projected_2024}
J.~K. Christopher, S.~Baek, and F.~Fioretto.
\newblock Projected {Generative} {Diffusion} {Models} for {Constraint} {Satisfaction}, Feb. 2024.
\newblock URL \url{http://arxiv.org/abs/2402.03559}.
\newblock arXiv:2402.03559 [cs].

\bibitem[Cicek et~al.(2016)Cicek, Abdulkadir, Lienkamp, Brox, and Ronneberger]{cicek_3d_2016}
O.~Cicek, A.~Abdulkadir, S.~S. Lienkamp, T.~Brox, and O.~Ronneberger.
\newblock {3D} {U}-{Net}: {Learning} {Dense} {Volumetric} {Segmentation} from {Sparse} {Annotation}, June 2016.
\newblock URL \url{http://arxiv.org/abs/1606.06650}.
\newblock arXiv:1606.06650 [cs].

\bibitem[Ding et~al.(2024)Ding, Zhang, Tian, and Zheng]{ding_diffusion_2024}
Z.~Ding, A.~Zhang, Y.~Tian, and Q.~Zheng.
\newblock Diffusion {World} {Model}, Feb. 2024.
\newblock URL \url{http://arxiv.org/abs/2402.03570}.
\newblock arXiv:2402.03570 [cs].

\bibitem[Durall et~al.(2023)Durall, Ghanim, Fernandez, Ettrich, and Keuper]{durall_deep_2023}
R.~Durall, A.~Ghanim, M.~R. Fernandez, N.~Ettrich, and J.~Keuper.
\newblock Deep diffusion models for seismic processing.
\newblock \emph{Computers \& Geosciences}, 177:\penalty0 105377, Aug. 2023.
\newblock ISSN 0098-3004.
\newblock \doi{10.1016/j.cageo.2023.105377}.
\newblock URL \url{https://www.sciencedirect.com/science/article/pii/ S009830042300081X}.

\bibitem[Finzi et~al.(2023)Finzi, Boral, Wilson, Sha, and Zepeda-Núñez]{finzi_user-defined_2023}
M.~Finzi, A.~Boral, A.~G. Wilson, F.~Sha, and L.~Zepeda-Núñez.
\newblock User-defined {Event} {Sampling} and {Uncertainty} {Quantification} in {Diffusion} {Models} for {Physical} {Dynamical} {Systems}, June 2023.
\newblock URL \url{http://arxiv.org/abs/2306.07526}.
\newblock arXiv:2306.07526 [cs].

\bibitem[Fort et~al.(2020)Fort, Hu, and Lakshminarayanan]{fort_deep_2020}
S.~Fort, H.~Hu, and B.~Lakshminarayanan.
\newblock Deep {Ensembles}: {A} {Loss} {Landscape} {Perspective}, June 2020.
\newblock URL \url{http://arxiv.org/abs/1912.02757}.
\newblock arXiv:1912.02757 [cs, stat].

\bibitem[Gahlot et~al.(2023)Gahlot, Erdinc, Orozco, Yin, and Herrmann]{gahlot_inference_2023}
A.~P. Gahlot, H.~T. Erdinc, R.~Orozco, Z.~Yin, and F.~J. Herrmann.
\newblock Inference of {CO2} flow patterns -- a feasibility study, Nov. 2023.
\newblock URL \url{http://arxiv.org/abs/2311.00290}.
\newblock arXiv:2311.00290 [math-ph, physics:physics].

\bibitem[Goodfellow et~al.(2014)Goodfellow, Pouget-Abadie, Mirza, Xu, Warde-Farley, Ozair, Courville, and Bengio]{goodfellow_generative_2014}
I.~Goodfellow, J.~Pouget-Abadie, M.~Mirza, B.~Xu, D.~Warde-Farley, S.~Ozair, A.~Courville, and Y.~Bengio.
\newblock Generative {Adversarial} {Nets}.
\newblock In \emph{Advances in {Neural} {Information} {Processing} {Systems}}, volume~27. Curran Associates, Inc., 2014.
\newblock URL \url{https://proceedings.neurips.cc/paper_files/paper/2014/hash/ 5ca3e9b122f61f8f06494c97b1afccf3-Abstract.html}.

\bibitem[Graupner et~al.(2011)Graupner, Li, and Bauer]{graupner_coupled_2011}
B.~J. Graupner, D.~Li, and S.~Bauer.
\newblock The coupled simulator {ECLIPSE}–{OpenGeoSys} for the simulation of {CO2} storage in saline formations.
\newblock \emph{Energy Procedia}, 4:\penalty0 3794--3800, Jan. 2011.
\newblock ISSN 1876-6102.
\newblock \doi{10.1016/j.egypro.2011.02.314}.
\newblock URL \url{https://www.sciencedirect.com/science/article/pii/ S1876610211005935}.

\bibitem[Hicks et~al.(2016)Hicks, Hoeber, Houbiers, Lescoffit, Ratcliffe, and Vinje]{hicks_time-lapse_2016}
E.~Hicks, H.~Hoeber, M.~Houbiers, S.~P. Lescoffit, A.~Ratcliffe, and V.~Vinje.
\newblock Time-lapse full-waveform inversion as a reservoir-monitoring tool — {A} {North} {Sea} case study.
\newblock \emph{The Leading Edge}, 35\penalty0 (10):\penalty0 850--858, Oct. 2016.
\newblock ISSN 1070-485X.
\newblock \doi{10.1190/tle35100850.1}.
\newblock URL \url{https://library.seg.org/doi/abs/10.1190/tle35100850.1}.
\newblock Publisher: Society of Exploration Geophysicists.

\bibitem[Ho and Salimans(2021)]{ho_classier-free_2021}
J.~Ho and T.~Salimans.
\newblock Classiﬁer-{Free} {Diffusion} {Guidance}.
\newblock \emph{NeurIPS 2021 Workshop on Deep Generative Models and Downstream Applications}, 2021.

\bibitem[Ho et~al.(2020)Ho, Jain, and Abbeel]{ho_denoising_2020}
J.~Ho, A.~Jain, and P.~Abbeel.
\newblock Denoising {Diffusion} {Probabilistic} {Models}.
\newblock In \emph{34th {Conference} on {Neural} {Information} {Processing} {Systems}}, page~25, 2020.

\bibitem[Ho et~al.(2022)Ho, Salimans, Gritsenko, Chan, Norouzi, and Fleet]{ho_video_2022}
J.~Ho, T.~Salimans, A.~Gritsenko, W.~Chan, M.~Norouzi, and D.~J. Fleet.
\newblock Video {Diffusion} {Models}, June 2022.
\newblock URL \url{http://arxiv.org/abs/2204.03458}.
\newblock arXiv:2204.03458 [cs].

\bibitem[Hu et~al.(2023)Hu, Grana, and Innanen]{hu_feasibility_2023}
Q.~Hu, D.~Grana, and K.~A. Innanen.
\newblock Feasibility of seismic time-lapse monitoring of {CO2} with rock physics parametrized full waveform inversion.
\newblock \emph{Geophysical Journal International}, 233\penalty0 (1):\penalty0 402--419, Apr. 2023.
\newblock ISSN 0956-540X.
\newblock \doi{10.1093/gji/ggac462}.
\newblock URL \url{https://doi.org/10.1093/gji/ggac462}.

\bibitem[Huang(2025)]{Huang2024diffusion-dataset}
X.~Huang.
\newblock Diffusion-based subsurface co\$\_2\$ multiphysics monitoring and forecasting.
\newblock [Dataset], Jan. 2025.
\newblock URL \url{https://doi.org/10.5281/zenodo.14681997}.

\bibitem[Huang and Alkhalifah(2024)]{huang_learned_2024}
X.~Huang and T.~Alkhalifah.
\newblock Learned frequency-domain scattered wavefield solutions using neural operators, May 2024.
\newblock URL \url{http://arxiv.org/abs/2405.01272}.
\newblock arXiv:2405.01272 [physics].

\bibitem[Huang et~al.(2024)Huang, Shi, Gao, Wei, Zhang, Bian, Yang, and Liu]{huang_lordnet_2024}
X.~Huang, W.~Shi, X.~Gao, X.~Wei, J.~Zhang, J.~Bian, M.~Yang, and T.-Y. Liu.
\newblock {LordNet}: {An} efficient neural network for learning to solve parametric partial differential equations without simulated data.
\newblock \emph{Neural Networks}, 176:\penalty0 106354, Aug. 2024.
\newblock ISSN 0893-6080.
\newblock \doi{10.1016/j.neunet.2024.106354}.
\newblock URL \url{https://www.sciencedirect.com/science/article/pii/ S0893608024002788}.

\bibitem[IPCC(2022)]{ipcc_climate_2022}
IPCC.
\newblock \emph{Climate {Change} 2022: {Mitigation} of {Climate} {Change}. {Contribution} of {Working} {Group} {III} to the {Sixth} {Assessment} {Report} of the {Intergovernmental} {Panel} on {Climate} {Change}}.
\newblock Cambridge University Press., 2022.

\bibitem[Jones et~al.(2012)Jones, Edgar, Selvage, and Crook]{jones_building_2012}
C.~E. Jones, J.~A. Edgar, J.~I. Selvage, and H.~Crook.
\newblock Building {Complex} {Synthetic} {Models} to {Evaluate} {Acquisition} {Geometries} and {Velocity} {Inversion} {Technologies}.
\newblock In \emph{74th {EAGE} {Annual} {Conference} \& {Exhibition}}, page~cp. European Association of Geoscientists \& Engineers, June 2012.
\newblock ISBN 978-90-73834-27-9.
\newblock \doi{10.3997/2214-4609.20148575}.
\newblock URL \url{https://www.earthdoc.org/content/papers/10.3997/2214- 4609.20148575}.
\newblock ISSN: 2214-4609.

\bibitem[Kingma and Welling(2013)]{kingma_auto-encoding_2013}
D.~P. Kingma and M.~Welling.
\newblock Auto-{Encoding} {Variational} {Bayes}, 2013.
\newblock URL \url{http://arxiv.org/abs/1312.6114}.
\newblock arXiv:1312.6114 [cs, stat].

\bibitem[Li and Li(2021)]{li_neural_2021}
B.~Li and Y.~E. Li.
\newblock Neural {Network}-{Based} {CO2} {Interpretation} {From} {4D} {Sleipner} {Seismic} {Images}.
\newblock \emph{Journal of Geophysical Research: Solid Earth}, 126\penalty0 (12):\penalty0 e2021JB022524, 2021.
\newblock ISSN 2169-9356.
\newblock \doi{10.1029/2021JB022524}.
\newblock URL \url{https://onlinelibrary.wiley.com/doi/abs/10.1029/ 2021JB022524}.
\newblock \_eprint: https://onlinelibrary.wiley.com/doi/pdf/10.1029/2021JB022524.

\bibitem[Li et~al.(2018)Li, Zhou, Li, Duguid, Que, Xue, and Tan]{li_prediction_2018}
B.~Li, F.~Zhou, H.~Li, A.~Duguid, L.~Que, Y.~Xue, and Y.~Tan.
\newblock Prediction of {CO2} leakage risk for wells in carbon sequestration fields with an optimal artificial neural network.
\newblock \emph{International Journal of Greenhouse Gas Control}, 68:\penalty0 276--286, Jan. 2018.
\newblock ISSN 1750-5836.
\newblock \doi{10.1016/j.ijggc.2017.11.004}.
\newblock URL \url{https://www.sciencedirect.com/science/article/pii/ S1750583617303237}.

\bibitem[Li et~al.(2020{\natexlab{a}})Li, Xu, Harris, and Darve]{li_coupled_2020}
D.~Li, K.~Xu, J.~M. Harris, and E.~Darve.
\newblock Coupled {Time}-{Lapse} {Full}-{Waveform} {Inversion} for {Subsurface} {Flow} {Problems} {Using} {Intrusive} {Automatic} {Differentiation}.
\newblock \emph{Water Resources Research}, 56\penalty0 (8):\penalty0 e2019WR027032, 2020{\natexlab{a}}.
\newblock ISSN 1944-7973.
\newblock \doi{10.1029/2019WR027032}.
\newblock URL \url{https://onlinelibrary.wiley.com/doi/abs/10.1029/ 2019WR027032}.
\newblock \_eprint: https://onlinelibrary.wiley.com/doi/pdf/10.1029/2019WR027032.

\bibitem[Li and Alkhalifah(2022)]{li_target-oriented_2022}
Y.~Li and T.~Alkhalifah.
\newblock Target-{Oriented} {Time}-{Lapse} {Elastic} {Full}-{Waveform} {Inversion} {Constrained} by {Deep} {Learning}-{Based} {Prior} {Model}.
\newblock \emph{IEEE Transactions on Geoscience and Remote Sensing}, 60:\penalty0 1--12, 2022.
\newblock ISSN 0196-2892, 1558-0644.
\newblock \doi{10.1109/TGRS.2022.3186028}.
\newblock URL \url{https://ieeexplore.ieee.org/document/9808328/}.

\bibitem[Li et~al.(2020{\natexlab{b}})Li, Kovachki, Azizzadenesheli, Liu, Bhattacharya, Stuart, and Anandkumar]{li_fourier_2020}
Z.~Li, N.~Kovachki, K.~Azizzadenesheli, B.~Liu, K.~Bhattacharya, A.~Stuart, and A.~Anandkumar.
\newblock Fourier {Neural} {Operator} for {Parametric} {Partial} {Differential} {Equations}, 2020{\natexlab{b}}.
\newblock URL \url{http://arxiv.org/abs/2010.08895}.
\newblock arXiv: 2010.08895.

\bibitem[Louboutin et~al.(2023{\natexlab{a}})Louboutin, Witte, Yin, Modzelewski, Kerim, Costa, and Nogueira]{louboutin_slimgroupjudijl_2023}
M.~Louboutin, P.~Witte, Z.~Yin, H.~Modzelewski, Kerim, C.~d. Costa, and P.~Nogueira.
\newblock slimgroup/{JUDI}.jl: v3.2.3.
\newblock [Software], Mar. 2023{\natexlab{a}}.
\newblock URL \url{https://zenodo.org/records/7785440}.

\bibitem[Louboutin et~al.(2023{\natexlab{b}})Louboutin, Yin, Orozco, Grady, Siahkoohi, Rizzuti, Witte, Møyner, Gorman, and Herrmann]{louboutin_learned_2023}
M.~Louboutin, Z.~Yin, R.~Orozco, T.~J. Grady, A.~Siahkoohi, G.~Rizzuti, P.~A. Witte, O.~Møyner, G.~J. Gorman, and F.~J. Herrmann.
\newblock Learned multiphysics inversion with differentiable programming and machine learning.
\newblock \emph{The Leading Edge}, 42\penalty0 (7):\penalty0 474--486, July 2023{\natexlab{b}}.
\newblock ISSN 1070-485X, 1938-3789.
\newblock \doi{10.1190/tle42070474.1}.
\newblock URL \url{https://library.seg.org/doi/10.1190/tle42070474.1}.

\bibitem[Lumley(2010)]{lumley_4d_2010}
D.~Lumley.
\newblock {4D} seismic monitoring of {CO2} sequestration.
\newblock \emph{The Leading Edge}, 29\penalty0 (2):\penalty0 150--155, Feb. 2010.
\newblock ISSN 1070-485X, 1938-3789.
\newblock \doi{10.1190/1.3304817}.
\newblock URL \url{https://library.seg.org/doi/10.1190/1.3304817}.

\bibitem[Micikevicius et~al.(2018)Micikevicius, Narang, Alben, Diamos, Elsen, Garcia, Ginsburg, Houston, Kuchaiev, Venkatesh, and Wu]{micikevicius_mixed_2018}
P.~Micikevicius, S.~Narang, J.~Alben, G.~Diamos, E.~Elsen, D.~Garcia, B.~Ginsburg, M.~Houston, O.~Kuchaiev, G.~Venkatesh, and H.~Wu.
\newblock Mixed {Precision} {Training}, Feb. 2018.
\newblock URL \url{http://arxiv.org/abs/1710.03740}.
\newblock arXiv:1710.03740 [cs, stat].

\bibitem[Møyner et~al.(2023)Møyner, Johnsrud, Nilsen, Raynaud, Lye, and Yin]{moyner_sintefmathjutuljl_2023}
O.~Møyner, M.~Johnsrud, H.~M. Nilsen, X.~Raynaud, K.~O. Lye, and Z.~Yin.
\newblock sintefmath/{Jutul}.jl: v0.2.6.
\newblock [Software], Apr. 2023.
\newblock URL \url{https://zenodo.org/records/7855605}.

\bibitem[Nichol and Dhariwal(2021)]{nichol_improved_2021}
A.~Q. Nichol and P.~Dhariwal.
\newblock Improved {Denoising} {Diffusion} {Probabilistic} {Models}.
\newblock In \emph{Proceedings of the 38th {International} {Conference} on {Machine} {Learning}}, pages 8162--8171. PMLR, July 2021.
\newblock URL \url{https://proceedings.mlr.press/v139/nichol21a.html}.
\newblock ISSN: 2640-3498.

\bibitem[Peebles and Xie(2022)]{peebles_scalable_2022}
W.~Peebles and S.~Xie.
\newblock Scalable {Diffusion} {Models} with {Transformers}, Dec. 2022.
\newblock URL \url{http://arxiv.org/abs/2212.09748}.
\newblock arXiv:2212.09748 [cs].

\bibitem[Price et~al.(2023)Price, Sanchez-Gonzalez, Alet, Ewalds, El-Kadi, Stott, Mohamed, Battaglia, Lam, and Willson]{price_gencast_2023}
I.~Price, A.~Sanchez-Gonzalez, F.~Alet, T.~Ewalds, A.~El-Kadi, J.~Stott, S.~Mohamed, P.~Battaglia, R.~Lam, and M.~Willson.
\newblock {GenCast}: {Diffusion}-based ensemble forecasting for medium-range weather, Dec. 2023.
\newblock URL \url{http://arxiv.org/abs/2312.15796}.
\newblock arXiv:2312.15796 [physics].

\bibitem[Pruess and Nordbotten(2011)]{pruess_numerical_2011}
K.~Pruess and J.~Nordbotten.
\newblock Numerical {Simulation} {Studies} of the {Long}-term {Evolution} of a {CO2} {Plume} in a {Saline} {Aquifer} with a {Sloping} {Caprock}.
\newblock \emph{Transport in Porous Media}, 90\penalty0 (1):\penalty0 135--151, Oct. 2011.
\newblock ISSN 0169-3913, 1573-1634.
\newblock \doi{10.1007/s11242-011-9729-6}.
\newblock URL \url{http://link.springer.com/10.1007/s11242-011-9729-6}.

\bibitem[Rezende and Mohamed(2015)]{rezende_variational_2015}
D.~Rezende and S.~Mohamed.
\newblock Variational {Inference} with {Normalizing} {Flows}.
\newblock In \emph{Proceedings of the 32nd {International} {Conference} on {Machine} {Learning}}, pages 1530--1538. PMLR, June 2015.
\newblock URL \url{https://proceedings.mlr.press/v37/rezende15.html}.
\newblock ISSN: 1938-7228.

\bibitem[Richardson(2023)]{richardson_deepwave_2023}
A.~Richardson.
\newblock Deepwave.
\newblock [Software], Sept. 2023.
\newblock URL \url{https://zenodo.org/records/8381177}.

\bibitem[Ringrose(2020)]{ringrose_co2_2020}
P.~Ringrose.
\newblock {CO2} {Storage} {Project} {Design}.
\newblock In P.~Ringrose, editor, \emph{How to {Store} {CO2} {Underground}: {Insights} from early-mover {CCS} {Projects}}, pages 85--126. Springer International Publishing, Cham, 2020.
\newblock ISBN 978-3-030-33113-9.
\newblock \doi{10.1007/978-3-030-33113-9_3}.
\newblock URL \url{https://doi.org/10.1007/978-3-030-33113-9_3}.

\bibitem[Rombach et~al.(2022)Rombach, Blattmann, Lorenz, Esser, and Ommer]{rombach_high-resolution_2022}
R.~Rombach, A.~Blattmann, D.~Lorenz, P.~Esser, and B.~Ommer.
\newblock High-{Resolution} {Image} {Synthesis} with {Latent} {Diffusion} {Models}, Apr. 2022.
\newblock URL \url{http://arxiv.org/abs/2112.10752}.
\newblock arXiv:2112.10752 [cs].

\bibitem[Shaw et~al.(2018)Shaw, Uszkoreit, and Vaswani]{shaw_self-attention_2018}
P.~Shaw, J.~Uszkoreit, and A.~Vaswani.
\newblock Self-{Attention} with {Relative} {Position} {Representations}, Apr. 2018.
\newblock URL \url{http://arxiv.org/abs/1803.02155}.
\newblock arXiv:1803.02155 [cs].

\bibitem[Song et~al.(2022)Song, Meng, and Ermon]{song_denoising_2022}
J.~Song, C.~Meng, and S.~Ermon.
\newblock Denoising {Diffusion} {Implicit} {Models}, Oct. 2022.
\newblock URL \url{http://arxiv.org/abs/2010.02502}.
\newblock arXiv:2010.02502 [cs].

\bibitem[Song et~al.(2023)Song, Dhariwal, Chen, and Sutskever]{song_consistency_2023}
Y.~Song, P.~Dhariwal, M.~Chen, and I.~Sutskever.
\newblock Consistency {Models}, Mar. 2023.
\newblock URL \url{http://arxiv.org/abs/2303.01469}.
\newblock arXiv:2303.01469 [cs, stat].

\bibitem[Stepien et~al.(2023)Stepien, Ferreira, Hosseinzadehsadati, Kadeethum, and Nick]{stepien_continuous_2023}
M.~Stepien, C.~A. Ferreira, S.~Hosseinzadehsadati, T.~Kadeethum, and H.~M. Nick.
\newblock Continuous conditional generative adversarial networks for data-driven modelling of geologic {CO} 2 storage and plume evolution.
\newblock \emph{Gas Science and Engineering}, 115:\penalty0 204982, July 2023.
\newblock ISSN 29499089.
\newblock \doi{10.1016/j.jgsce.2023.204982}.
\newblock URL \url{https://linkinghub.elsevier.com/retrieve/pii/ S2949908923001103}.

\bibitem[Sun et~al.(2023)Sun, Leong, and Zhu]{sun_denoising_2023}
A.~Y. Sun, Z.~X. Leong, and T.~Zhu.
\newblock A denoising diffusion probabilistic modeling approach for predicting {CO} $_{\textrm{2}}$ plume evolution from seismic shot gathers.
\newblock In \emph{Third {International} {Meeting} for {Applied} {Geoscience} \& {Energy} {Expanded} {Abstracts}}, pages 376--380, Houston, Texas, Dec. 2023. Society of Exploration Geophysicists and American Association of Petroleum Geologists.
\newblock \doi{10.1190/image2023-3915826.1}.
\newblock URL \url{https://library.seg.org/doi/10.1190/image2023-3915826.1}.

\bibitem[Wang et~al.(2023)Wang, Huang, and Alkhalifah]{wang_prior_2023}
F.~Wang, X.~Huang, and T.~A. Alkhalifah.
\newblock A {Prior} {Regularized} {Full} {Waveform} {Inversion} {Using} {Generative} {Diffusion} {Models}.
\newblock \emph{IEEE Transactions on Geoscience and Remote Sensing}, 61:\penalty0 1--11, 2023.
\newblock ISSN 1558-0644.
\newblock \doi{10.1109/TGRS.2023.3337014}.
\newblock URL \url{https://ieeexplore.ieee.org/document/10328845}.
\newblock Conference Name: IEEE Transactions on Geoscience and Remote Sensing.

\bibitem[Wang et~al.(2024)Wang, Huang, and Alkhalifah]{wang_controllable_2024}
F.~Wang, X.~Huang, and T.~Alkhalifah.
\newblock Controllable {Seismic} {Velocity} {Synthesis} {Using} {Generative} {Diffusion} {Models}.
\newblock \emph{Journal of Geophysical Research: Machine Learning and Computation}, 1\penalty0 (3):\penalty0 e2024JH000153, 2024.
\newblock ISSN 2993-5210.
\newblock \doi{10.1029/2024JH000153}.
\newblock URL \url{https://onlinelibrary.wiley.com/doi/abs/10.1029/ 2024JH000153}.
\newblock \_eprint: https://onlinelibrary.wiley.com/doi/pdf/10.1029/2024JH000153.

\bibitem[Wen et~al.(2021)Wen, Hay, and Benson]{wen_ccsnet_2021}
G.~Wen, C.~Hay, and S.~M. Benson.
\newblock {CCSNet}: {A} deep learning modeling suite for {CO} 2 storage.
\newblock \emph{Advances in Water Resources}, 155:\penalty0 104009, Sept. 2021.
\newblock ISSN 03091708.
\newblock \doi{10.1016/j.advwatres.2021.104009}.
\newblock URL \url{https://linkinghub.elsevier.com/retrieve/pii/ S0309170821001640}.

\bibitem[Wen et~al.(2022)Wen, Li, Long, Azizzadenesheli, Anandkumar, and Benson]{wen_accelerating_2022}
G.~Wen, Z.~Li, Q.~Long, K.~Azizzadenesheli, A.~Anandkumar, and S.~M. Benson.
\newblock Accelerating {Carbon} {Capture} and {Storage} {Modeling} using {Fourier} {Neural} {Operators}, Oct. 2022.
\newblock URL \url{http://arxiv.org/abs/2210.17051}.
\newblock arXiv:2210.17051 [physics].

\bibitem[Wilson and Izmailov(2022)]{wilson_bayesian_2022}
A.~G. Wilson and P.~Izmailov.
\newblock Bayesian {Deep} {Learning} and a {Probabilistic} {Perspective} of {Generalization}, Mar. 2022.
\newblock URL \url{http://arxiv.org/abs/2002.08791}.
\newblock arXiv:2002.08791 [cs, stat].

\bibitem[Witte et~al.(2023)Witte, Konuk, Skjetne, and Chandra]{witte_fast_2023}
P.~A. Witte, T.~Konuk, E.~Skjetne, and R.~Chandra.
\newblock Fast {CO2} saturation simulations on large-scale geomodels with artificial intelligence-based {Wavelet} {Neural} {Operators}.
\newblock \emph{International Journal of Greenhouse Gas Control}, 126:\penalty0 103880, June 2023.
\newblock ISSN 17505836.
\newblock \doi{10.1016/j.ijggc.2023.103880}.
\newblock URL \url{https://linkinghub.elsevier.com/retrieve/pii/ S1750583623000506}.

\bibitem[Xu et~al.(2024)Xu, Mi, Wang, and Chen]{xu_towards_2024}
T.~Xu, P.~Mi, R.~Wang, and Y.~Chen.
\newblock Towards {Faster} {Training} of {Diffusion} {Models}: {An} {Inspiration} of {A} {Consistency} {Phenomenon}, Mar. 2024.
\newblock URL \url{http://arxiv.org/abs/2404.07946}.
\newblock arXiv:2404.07946 [cs].

\bibitem[Yin(2022)]{yin_slimgroupseis4ccsjl_2022}
Z.~Yin.
\newblock slimgroup/{Seis4CCS}.jl.
\newblock [Software], 2022.
\newblock URL \url{https://github.com/slimgroup/Seis4CCS.jl}.
\newblock original-date: 2021-06-02T16:32:38Z.

\bibitem[Yin et~al.(2023{\natexlab{a}})Yin, Bruer, and Louboutin]{yin_slimgroupjutuldarcyrulesjl_2023}
Z.~Yin, G.~Bruer, and M.~Louboutin.
\newblock slimgroup/{JutulDarcyRules}.jl: v0.2.5.
\newblock [Software], Apr. 2023{\natexlab{a}}.
\newblock URL \url{https://zenodo.org/records/7863970}.

\bibitem[Yin et~al.(2023{\natexlab{b}})Yin, Erdinc, Gahlot, Louboutin, and Herrmann]{yin_derisking_2023}
Z.~Yin, H.~T. Erdinc, A.~P. Gahlot, M.~Louboutin, and F.~J. Herrmann.
\newblock Derisking geologic carbon storage from high-resolution time-lapse seismic to explainable leakage detection.
\newblock \emph{The Leading Edge}, 42\penalty0 (1):\penalty0 69--76, Jan. 2023{\natexlab{b}}.
\newblock ISSN 1070-485X.
\newblock \doi{10.1190/tle42010069.1}.
\newblock URL \url{https://library.seg.org/doi/full/10.1190/tle42010069.1}.
\newblock Publisher: Society of Exploration Geophysicists.

\bibitem[Zeng and Wang(2012)]{zeng_3d-ssim_2012}
K.~Zeng and Z.~Wang.
\newblock {3D}-{SSIM} for video quality assessment.
\newblock In \emph{2012 19th {IEEE} {International} {Conference} on {Image} {Processing}}, pages 621--624, Sept. 2012.
\newblock \doi{10.1109/ICIP.2012.6466936}.
\newblock URL \url{https://ieeexplore.ieee.org/document/6466936}.
\newblock ISSN: 2381-8549.

\bibitem[Zhang et~al.(2024)Zhang, Li, and Huang]{zhang_conditional_2024}
H.~Zhang, Y.~Li, and J.~Huang.
\newblock Conditional {Denoising} {Diffusion} {Probabilistic} {Model} for {Seismic} {Diffraction} {Separation} and {Imaging}.
\newblock \emph{IEEE Transactions on Geoscience and Remote Sensing}, 62:\penalty0 1--13, 2024.
\newblock ISSN 1558-0644.
\newblock \doi{10.1109/TGRS.2024.3381193}.
\newblock URL \url{https://ieeexplore.ieee.org/abstract/document/10478607}.
\newblock Conference Name: IEEE Transactions on Geoscience and Remote Sensing.

\bibitem[Zhong et~al.(2019)Zhong, Sun, and Jeong]{zhong_predicting_2019}
Z.~Zhong, A.~Y. Sun, and H.~Jeong.
\newblock Predicting {CO2} {Plume} {Migration} in {Heterogeneous} {Formations} {Using} {Conditional} {Deep} {Convolutional} {Generative} {Adversarial} {Network}.
\newblock \emph{Water Resources Research}, 55\penalty0 (7):\penalty0 5830--5851, 2019.
\newblock ISSN 1944-7973.
\newblock \doi{10.1029/2018WR024592}.
\newblock URL \url{https://onlinelibrary.wiley.com/doi/abs/10.1029/ 2018WR024592}.
\newblock \_eprint: https://onlinelibrary.wiley.com/doi/pdf/10.1029/2018WR024592.

\bibitem[Zhou et~al.(2019)Zhou, Lin, Zhang, Wu, Wang, Dilmore, and Guthrie]{zhou_data-driven_2019}
Z.~Zhou, Y.~Lin, Z.~Zhang, Y.~Wu, Z.~Wang, R.~Dilmore, and G.~Guthrie.
\newblock A data-driven {CO2} leakage detection using seismic data and spatial–temporal densely connected convolutional neural networks.
\newblock \emph{International Journal of Greenhouse Gas Control}, 90:\penalty0 102790, Nov. 2019.
\newblock ISSN 1750-5836.
\newblock \doi{10.1016/j.ijggc.2019.102790}.
\newblock URL \url{https://www.sciencedirect.com/science/article/pii/ S1750583619301239}.

\end{thebibliography}
\end{document}